\documentclass[12pt,twoside,a4paper]{article}
\pdfoutput=1

\usepackage{amsmath, amsthm, amsfonts, amssymb}
\usepackage{graphicx}
\usepackage{float}
\usepackage{color}
\usepackage{cite}
\usepackage[width=\textwidth,font=small,labelfont=bf,
labelsep=endash]{caption}
\usepackage{titlesec}

\titleformat{\section}
  {\normalfont\fontsize{13}{16}\bfseries}{\thesection}{1em}{}
\titleformat{\subsection}
  {\normalfont\fontsize{11}{14}\bfseries}{\thesubsection}{1em}{}

\normalsize
\newcommand\Lame {Lam\'e\ }
\newcommand\Backlund {B\"{a}cklund }
\newcommand\Schrodinger {Schr\"{o}dinger }

\newcommand\tr {\mathrm{Tr}}
\newcommand\diag {\mathrm{diag}}
\newcommand\sign {\mathrm{sgn}}

\newcommand\csch {\mathrm{csch}}

\renewcommand{\theequation}{\arabic{section}.\arabic{equation}}

\begin{document}

\title{\textbf{Dressed Elliptic String Solutions on $\mathbb{R}\times$S$^2$}}
\author{Dimitrios Katsinis$^{1,2}$, Ioannis Mitsoulas$^3$ and Georgios Pastras$^2$}
\date{\small $^1$Department of Physics, National and Kapodistrian University of Athens,\\University Campus, Zografou, Athens 15784, Greece\\
$^2$NCSR ``Demokritos'', Institute of Nuclear and Particle Physics,\\Aghia Paraskevi 15310, Attiki, Greece\\
$^3$Department of Physics, School of Applied Mathematics and Physical Sciences,\\National Technical University, Athens 15780, Greece\linebreak \vspace{8pt}
\texttt{dkatsinis@phys.uoa.gr, mitsoula@central.ntua.gr, pastras@inp.demokritos.gr}}

\vskip .5cm

\maketitle

\begin{abstract}
We obtain classical string solutions on $\mathbb{R}^t \times $S$^2$ by applying the dressing method on string solutions with elliptic Pohlmeyer counterparts. This is realized through the use of the simplest possible dressing factor, which possesses just a pair of poles lying on the unit circle. The latter is equivalent to the action of a single \Backlund transformation on the corresponding sine-Gordon solutions. The obtained dressed elliptic strings present an interesting bifurcation of their qualitative characteristics at a specific value of a modulus of the seed solutions. Finally, an interesting generic feature of the dressed strings, which originates from the form of the simplest dressing factor and not from the specific seed solution, is the fact that they can be considered as drawn by an epicycle of constant radius whose center is running on the seed solution. The radius of the epicycle is directly related to the location of the poles of the dressing factor. \newline \newline \textbf{Keywords:} Classical Strings, Integrable Systems, Dressing Method, Pohlmeyer Reduction
\end{abstract}

\newpage

\tableofcontents

\newpage

\setcounter{equation}{0}
\section{Introduction}
\label{sec:introduction}

The holographic duality AdS/CFT \cite{ads_malda,ads_GKP,ads_witten} forms a broad framework, which connects gravitational theories in spaces with AdS asymptotics to conformal field theories defined on their respective boundary. As a weak/strong duality, it links the strongly (weakly) coupled regime of any one of the two theories to the weakly (strongly) coupled regime of its dual counterpart. The holographic duality has found many applications on both sides of it, such as in the study of strongly coupled CFTs (hydrodynamics, condensed matter systems and so on), and in the study of strongly coupled gravitational dynamics.

Classical string solutions have shed light to many aspects of the holographic duality. Such solutions correspond to the planar limit, where the rank of the gauge group of the boundary theory is infinite keeping the t'Hooft coupling finite but large enough in order to neglect the backreaction of the string to the background geometry. Thus, they probe non-perturbative effects of a large $N$ boundary CFT. A wide class of such solutions propagating on the sphere, on AdS space or on their tensor product has been found in the literature. Such solutions include the GKP string \cite{GKP_string}, the BMN particle \cite{BMN}, the giant magnons \cite{Giant_Magnons,Dyonic_Giant_Magnons}, the single spikes \cite{single_spike_rs2} as well as a wider class of spiky string solutions \cite{dual_spikes,spiky_string,multi,helical,Kruczenski:2006pk}, which includes the former as special limits \cite{part1}. (See also \cite{tseytlin_review}, for a review of the subject.) 

The non-linear sigma models (NLSMs) that describe strings propagating in symmetric spaces, are well known to be reducible to integrable systems of the same family as the sine-Gordon equation and multi-component generalizations of the latter \cite{Barbashov:1980kz,DeVega:1992xc,Larsen:1996gn,Grigoriev:2007bu}. This procedure, widely known as Pohlmeyer reduction \cite{Pohlmeyer:1975nb,Zakharov:1973pp} is non-trivial, since the transformation connecting the original NLSM fields to the field variables of the reduced theory is non-local. Despite this fact, it has be shown that the reduced system can also be derived from a local Lagrangian being a gauged WZW model with an integrable potential \cite{Bakas:1993xh,Bakas:1995bm,FernandezPousa:1996hi,Miramontes:2008wt}.

The integrable systems of the family of the sine-Gordon equation possess \Backlund transformations, which connect solutions in pairs. Given a seed solution, these transformations generate a new non-trivial one. Iterative application of the \Backlund transformations leads to infinite towers of solutions. The archetypical example is the sine-Gordon equation, where using the vacuum as seed solution, one can build the one-kink solutions and then a tower of multi-kink/breather solutions \cite{Faddeev:1974em}. The analogue of this procedure in the NLSM is the so called ``dressing method'' \cite{Harnad:1983we,Pohl_avatars}. This method has been applied in the literature to produce string solutions on dS space \cite{Combes:1993rw}, on the sphere \cite{Spradlin:2006wk,Kalousios:2006xy} and on AdS space \cite{Jevicki:2007pk,Jevicki:2007aa} that correspond to one- or multi-kink solutions of the Pohlmeyer reduced system.

Although the procedure of Pohlmeyer reduction is straightforward, in other words it is trivial to find the solution of the Pohlmeyer reduced system, given a solution of the NLSM, the inverse is highly non-trivial for two reasons: firstly due to the fact that Pohlmeyer reduction involves a non-local transformation and secondly due to the fact that Pohlmeyer reduction is a many-to-one mapping; there are many NLSM solutions with the same Pohlmeyer counterpart. For this reason, the accumulated knowledge about the integrable sine-Gordon systems is ineffective in the NLSM case. Nevertheless, recently, a method for the inversion of Pohlmeyer reduction was developed \cite{bakas_pastras,Pastras:2016vqu}, which can be applied in the case of elliptic solutions of the reduced system. This method implements a connection between solutions of the NLSM and the eigenfunctions of the $n=1$ \Lame problem in order to construct the NLSM solutions with elliptic Pohlmeyer counterparts. In the case of strings propagating on $\mathbb{R}^t\times$S$^2$ \cite{part1}, it turns out that these are the spiky strings and their various limits.

In this work, we use classical elliptic string solutions as seed for the construction of higher genus string solutions on $\mathbb{R}^t\times$S$^2$, via the dressing method. This is made possible due to the simple and universal description of the elliptic solutions achieved in our previous work \cite{part1} via the inversion of Pohlmeyer reduction. We carry out this study in both the NLSM and the Pohlmeyer reduced theory, namely the sine-Gordon equation, in order to understand the correspondence between the dressing method and the \Backlund transformations of the latter more deeply. 

Although more general higher genus solutions of both the NLSM and the sine-Gordon equation can be expressed in terms of Riemann's hyperelliptic theta function \cite{Kotlarov:2014yxa,Kozel,Dorey:2006zj}, it is difficult to study their properties in this form. Unlike this approach, the solutions presented in this work are genus two solutions, which are expressed in terms of simple trigonometric and elliptic functions, and, thus, their properties can be studied analytically. This study is the first application of the dressing method on a non-trivial background, whose Pohlmeyer counterpart is neither the vacuum nor a kink solution, i.e. a solution connected to the vacuum via \Backlund transformations \cite{Spradlin:2006wk,Kalousios:2006xy}. The development of this kind of solutions can also be very useful in systems whose Pohlmeyer reduced theory does not possess a vacuum solution; the cosh-Gordon equation is such an example \cite{Pastras:2016vqu}.

The structure of the paper is as follows: In section \ref{sec:review} we review the construction of the elliptic string solutions on $\mathbb{R}^t\times$S$^2$ presented in \cite{part1}, as well as their Pohlmeyer counterparts. In section \ref{sec:dress_review}, we review the dressing method and in section \ref{sec:dressed_strings}, we apply it to obtain the dressed elliptic string solutions. In section \ref{sec:Elliptic_kinks}, we study the relation between the dressing method and the \Backlund transformations of the sine-Gordon equation and we obtain the Pohlmeyer counterparts of the dressed elliptic string solutions presented in section \ref{sec:dressed_strings}. In section \ref{sec:discussion} we discuss our results and possible future extensions. Finally, there is an appendix containing some interesting limits of the sine-Gordon solutions and some more technical details on the dressing method. 

\setcounter{equation}{0}
\section{Review of Elliptic String Solutions on $\mathbb{R}^t \times$S$^2$}
\label{sec:review}
The non-linear sigma models describing strings propagating in symmetric spaces are reducible to integrable systems similar to the sine-Gordon equation, a procedure widely known as Pohlmeyer reduction. A typical example is that of strings propagating on $\mathbb{R}^t \times$S$^2$ ($\mathbb{R}^t$ stands for time), which are reducible to the sine-Gordon equation itself. An important ingredient of the Pohlmeyer reduction is the embedding of the symmetric target space in a higher dimensional flat space, in this case the four-dimensional Minkowski space. In this language, the string action is written as
\begin{equation}
S = \int d\xi^+ d\xi^- \left( \left( \partial_+ X \right)\cdot \left( \partial_- X \right) + \lambda \left( \vec X \cdot \vec X - R^2 \right) \right) ,
\label{eq:Pohlmeyer_action}
\end{equation}
where $X \in \mathbb{R}^{(1,3)}$ and $\xi_\pm \equiv \left( \xi^1 \pm \xi^0 \right) / 2$. $A \cdot B$ stands for the inner product of two four-vectors $A$ and $B$ with respect to the Minkowski metric, $g = \diag \{ -1 ,1 ,1 ,1 \}$, while $\vec X$ stands for the three-vector composed by the three spatial components of $X$. The usual treatment of this system takes advantage of the $X^0$ equation of motion, $\partial_+\partial_- X^0 = 0$, to select a gauge (the static gauge), where the $X^0$ coordinate is proportional to the time-like worldsheet coordinate, namely $X^0 \sim \xi^0$. However, for our purposes, it is more suitable to select a more general gauge, which we will call the linear gauge, where it holds that
\begin{equation}
X^0 = m_+ \xi^+ + m_- \xi^- .
\end{equation}
Trivially, the linear and static gauges are connected via a boost in the worldsheet coordinates. In the linear gauge, Pohlmeyer reduction may be performed as usual, to show that the reduced system is the sine-Gordon equation
\begin{equation}
\partial_+ \partial_- \varphi = \mu^2 \sin \varphi ,
\label{eq:Pohlmeyer_SGequation}
\end{equation}
where $\mu^2 := - {m_+} {m_-} / {R^2}$ and the Pohlmeyer field $\varphi$ is defined as
\begin{equation}
{m_+} {m_-} \cos \varphi := \left( \partial_+ \vec X \right) \cdot \left( \partial_- \vec X \right) .
\label{eq:Pohlmeyer_sG_field_definition}
\end{equation}

The sine-Gordon equation has solutions which can be expressed in terms of elliptic functions and depend solely on either the time-like or the space-like worldsheet coordinate \cite{part1}. These read
\begin{equation}
\cos \varphi \left(\xi^0 , \xi^1; E \right) = \mp {\frac{1}{{{\mu ^2}}}\left( {2\wp \left( {\xi^{0/1} + {\omega _2}} \right) + \frac{E}{3}} \right)} ,
\label{eq:elliptic_solution}
\end{equation}
\pagebreak
which implies
\begin{equation}
\varphi \left(\xi^0 , \xi^1; E \right) = \begin{cases}
{\left( { - 1} \right)^{\left\lfloor {\frac{\xi^0}{{2{\omega _1}}}} \right\rfloor }}\arccos \left( - {\frac{{2\wp \left( {\xi^0 + {\omega _2}} \right) + \frac{E}{3}}}{{{\mu ^2}}}} \right), & E < \mu^2 , \\
{\left( { - 1} \right)^{\left\lfloor {\frac{\xi^0}{{{\omega _1}}}} \right\rfloor }}\arccos \left( - {\frac{{2\wp \left( {\xi^0 + {\omega _2}} \right) + \frac{E}{3}}}{{{\mu ^2}}}} \right) + 2\pi \left\lfloor {\frac{{\xi^0} + {\omega _1}}{{2{\omega _1}}}} \right\rfloor , & E > \mu^2 
\end{cases}
\label{eq:elliptic_solution_phi_0}
\end{equation}
for translationally invariant solutions and
\begin{equation}
\varphi \left(\xi^0 , \xi^1 ; E \right) = \begin{cases}
{\left( { - 1} \right)^{\left\lfloor {\frac{\xi^1}{{2{\omega _1}}}} \right\rfloor }}\arccos \left( - {\frac{{2\wp \left( {\xi^1 + {\omega _2}} \right) + \frac{E}{3}}}{{{\mu ^2}}}} \right) + \pi , & E < \mu^2 , \\
{\left( { - 1} \right)^{\left\lfloor {\frac{\xi^1}{{{\omega _1}}}} \right\rfloor }}\arccos \left( - {\frac{{2\wp \left( {\xi^1 + {\omega _2}} \right) + \frac{E}{3}}}{{{\mu ^2}}}} \right) + 2\pi \left\lfloor {\frac{{\xi^1} + {{\omega _1}}}{{2{\omega _1}}}} \right\rfloor + \pi , & E > \mu^2 
\end{cases}
\label{eq:elliptic_solution_phi_1}
\end{equation}
for the static ones. As equations \eqref{eq:elliptic_solution_phi_0} and \eqref{eq:elliptic_solution_phi_1} indicate, these elliptic solutions are identified by the value of the integration constant $E$, which may take any real value $E > -\mu^2$. In analogy to the simple pendulum, the elliptic solutions have different qualitative behaviour depending on whether the integration constant is smaller or larger than $\mu^2$; we will call the former as ``oscillatory'' solutions or trains of kink and anti-kink pairs and the latter as ``rotating'' solutions or trains of kinks.

In general, there is no systematic method to invert the Pohlmeyer reduction, as it was explained in the introduction. However, in the specific case of the elliptic solutions of the sine-Gordon equation, a systematic method to build the corresponding NLSM solutions has been developed \cite{part1}. This method was initially applied in the case of strings propagating in AdS$_3$ and dS$_3$ \cite{bakas_pastras} and subsequently for the construction of minimal surfaces in H$^3$ \cite{Pastras:2016vqu}. Given the specific solutions of the Pohlmeyer reduced system, the NLSM equations of motion can be solved via separation of variables, leading to pairs of effective \Schrodinger problems, each pair containing one flat potential and one $n=1$ \Lame potential. Using properties of the latter, it is possible to find appropriate solutions of the equations of motion that additionally satisfy the geometric and Virasoro constraints, effectively inverting the Pohlmeyer reduction for the class of elliptic solutions of the reduced system. The corresponding string solutions read
\begin{align}
t_{0/1} &= R\sqrt {{x_2} - \wp \left( a \right)} {\xi ^0} + R\sqrt {{x_3} - \wp \left( a \right)} {\xi ^1} , \label{eq:elliptic_solutions_t} \\
{\vec X}_{0/1} &= \frac{R}{\sqrt {{x_1} - \wp \left( a \right)}}\left( {\begin{array}{*{20}{c}}
{\sqrt {\wp \left( {{\xi ^{0/1}} + {\omega _2}} \right) - \wp \left( a \right)} \cos\left( {\ell {\xi ^{1/0}} - {\Phi }\left( {{\xi ^{0/1}};a} \right)} \right)}\\
{\sqrt {\wp \left( {{\xi ^{0/1}} + {\omega _2}} \right) - \wp \left( a \right)} \sin \left( {\ell {\xi ^{1/0}} - {\Phi }\left( {{\xi ^{0/1}};a} \right)} \right)}\\
{\sqrt {{x_1} - \wp \left( {{\xi ^{0/1}} + {\omega _2}} \right)} }
\end{array}} \right) , \label{eq:elliptic_solutions_review}
\end{align}
where the index $0/1$ denotes whether the Pohlmeyer counterpart of the solution is a translationally invariant or static solution of the sine-Gordon equation and the function $\Phi$ is defined as
\begin{equation}
{\Phi }\left( {{\xi };a} \right) :=  - \frac{i}{2}\ln \frac{{\sigma \left( {{\xi } + {\omega _2} + a} \right)\sigma \left( {{\omega _2} - a} \right)}}{{\sigma \left( {{\xi } + {\omega _2} - a} \right)\sigma \left( {{\omega _2} + a} \right)}} + i\zeta \left( a \right){\xi } .
\label{eq:elliptic_solutions_lame_phase_def}
\end{equation}
The function $\Phi$ is quasi-periodic, obeying
\begin{equation}
\Phi \left( {{\xi ^0} + 2{\omega _1}; a} \right) = \Phi \left( {{\xi ^0}; a} \right) + 2 i \left( {\zeta \left( { a} \right){\omega _1} - \zeta \left( {{\omega _1}} \right) a} \right) .
\label{eq:elliptic_solutions_lame_phase_quasiper}
\end{equation}
The moduli of the Weierstrass elliptic function in the above expressions are given by
\begin{equation}
g_2 = \frac{E^2}{3} + \mu^4 , \quad g_3 = \frac{E}{3} \left( \left( \frac{E}{3} \right)^2 - \mu^4 \right) ,
\end{equation}
the parameters $\ell$ and $\wp \left( a \right)$ are given by
\begin{equation}
\ell^2 = {{x_1} - \wp \left( a \right)} = \frac{{m_ + ^2 + m_ - ^2}}{{4{R^2}}} + \frac{{E}}{2}
\end{equation}
and $x_1$ is one of the roots of the cubic polynomial associated with the Weierstrass elliptic function, namely $x_1 = {E}/{3}$.

The parameter $a$ is a free parameter of the construction and can be selected anywhere in the imaginary axis. All solutions that correspond to the same parameter $E$ have the same Pohlmeyer counterpart, independently of the value of the parameter $a$ and form a Bonnet family of worldsheets. The sign of $a$ is connected to the sign of $\ell$  via
\begin{equation}
- i{c^2}\ell \wp '\left( a \right) = \frac{{m_ + ^2 - m_ - ^2}}{2} .
\label{eq:elliptic_solutions_sign_of_a}
\end{equation}

The solutions \eqref{eq:elliptic_solutions_review} form four classes of solutions, determined by whether the Pohlmeyer counterpart is oscillatory or rotating as well as static or translationally invariant. They are the known spiky/helical strings \cite{helical} and they have many interesting limits, such as the the GKP strings (static, $a=\omega_2$) \cite{GKP_string}, the BMN particle (translationally invariant, $E=-\mu^2$) \cite{BMN}, the giant magnons (static, $E=\mu^2$) \cite{Giant_Magnons} or the single spike (translationally invariant, $E=\mu^2$) \cite{single_spike_rs2}, which can be easily studied in this formulation.

In general, the elliptic string solutions in spherical coordinates can be written in the form
\begin{equation}
f\left( {\theta ,\varphi  - \omega t} \right) = 0 .
\end{equation}
where
\begin{equation}
\omega_{0/1}  = \frac{1}{R}\sqrt {\frac{{{x_1} - \wp \left( a \right)}}{{{x_{3/2}} - \wp \left( a \right)}}} .
\label{eq:elliptic_solutions_omega}
\end{equation}
The string solutions with static counterparts can be conceived as rigidly rotating string configurations. The ones with oscillatory counterparts are smooth, whereas the ones with rotating counterparts contain spikes, which move with the speed of light. Similarly, the string solutions with translationally invariant counterparts can be understood as wave propagating solutions and they are always spiky. 

In order to form a closed string of finite size, the parameter $a$ has to be selected, so that the string obeys appropriate periodic conditions. It turns out that the necessary condition is
\begin{equation}
i n_{0/1} \omega_1 \left( { \zeta \left( {{\omega _1}} \right)\frac{a}{\omega_1} + \zeta \left( {{\omega _{x_{3/2}}}} \right) - \zeta \left( a + {{\omega _{x_{3/2}}}} \right) } \right) = \pi ,
\end{equation}
where $n_{0/1}$ is an integer when the solution has a rotating counterpart and an even integer when it has an oscillatory counterpart. More information is provided in \cite{part1}.

\setcounter{equation}{0}
\section{Review of the Dressing Method}
\label{sec:dress_review}

The theories emerging after the Pohlmeyer reduction of the non-linear sigma models describing the propagation of classical strings in symmetric spaces possess \Backlund transformations, which connect pairs of solutions. These transformations are a manifestation of the model's integrability. The dressing method \cite{Zakharov:1973pp,Zakharov:1980ty,Harnad:1982cf,Harnad:1983we,Antoine:1986vf,Pohl_avatars,Combes:1993rw} is the direct analogue of the \Backlund transformations in the NLSM. In the literature, it has been used in order to generate non-trivial solutions \cite{Combes:1993rw,Spradlin:2006wk,Kalousios:2006xy,Jevicki:2007pk,Jevicki:2007aa}, whose seed solution corresponds to the vacuum of the reduced theory. In this section, we review a few elements of the theory of NLSMs on symmetric spaces, the dressing method in general, and the case of spheres S$^n$ in particular. This is by no means a complete review of the subject. It is rather a quick introduction to some concepts used in this paper. In the next section, we apply the dressing method on an elliptic seed string solution on S$^2$ in order to generate new non-trivial string solutions. In the following, without loss of generality, we take the radius of the target space sphere equal to one.

\subsection{The Non-linear Sigma Model}

The action of the non linear sigma model is
\begin{equation}
S=\frac{1}{8}\int\,d\xi_+d\xi_-\, \tr\left(\partial_+ f^{-1}\partial_- f\right) ,
\label{eq:dressing_review_nlsm_action}
\end{equation}
where $f$ takes values in the Lie group $F$ and it is a function of the worldsheet coordinates $\xi^\pm$. Varying this action with respect to $f$ yields the equation of motion
\begin{equation}
\partial_+\left(\partial_- ff^{-1}\right)+\partial_-\left(\partial_+ ff^{-1}\right)=0 .
\label{eq:dressing_review_eom}
\end{equation}

We introduce the currents $J_\pm := \partial_\pm ff^{-1}$, which allow the expression of the equation of motion \eqref{eq:dressing_review_eom} as
\begin{equation}
\partial_+  J_-+\partial_- J_+=0 .
\label{eq:dressing_review_eom_currents}
\end{equation}
By construction, the currents $J_\pm$ obey the relation
\begin{equation}
[\partial_+-J_+,\partial_--J_-]=0.
\label{eq:dressing_review_zero_curvature_currents}
\end{equation}
Introducing a complex parameter $\lambda$, equations \eqref{eq:dressing_review_eom_currents} and \eqref{eq:dressing_review_zero_curvature_currents} can be packed to one, namely,
\begin{equation}
\left[\partial_+-\frac{1}{1+\lambda} J_+ , \partial_--\frac{1}{1-\lambda} J_- \right] = 0 .
\label{eq:dressing_review_zero_curvature}
\end{equation}
In this form, equations \eqref{eq:dressing_review_eom_currents} and \eqref{eq:dressing_review_zero_curvature_currents} can be recovered as the residues of \eqref{eq:dressing_review_zero_curvature} at $\lambda = \pm 1$.
 
We introduce the following auxiliary system of first order differential equations
\begin{equation}
{\partial _ \pm }\Psi \left( \lambda  \right) = \frac{J_\pm}{{1 \pm \lambda }} \Psi \left( \lambda  \right) .
\label{eq:dressing_review_auxiliary}
\end{equation}
Equation \eqref{eq:dressing_review_zero_curvature} is just the compatibility condition for this system.

The NLSM action \eqref{eq:dressing_review_nlsm_action} is invariant under the transformations
 \begin{equation}
 f\to U_L \,f \,U_R,\quad U_{L,R}\in F .
 \end{equation}
Thus, it possesses a global $F_L\times F_R$ symmetry. The associated left and right conserved currents are
 \begin{equation}
 J^L_\mu=\partial_\mu ff^{-1} ,\quad J^R_\mu=f^{-1}\partial_\mu f ,
 \label{eq:dressing_review_currents_left_right}
 \end{equation}
respectively. Notice that the left current was already defined earlier, where we supressed the superscript $L$ for notational simplicity. In the following, we will continue to do so for the left currents and will only write the superscript $R$ for the right currents if necessary. The corresponding conserved charges are
 \begin{equation}
 \mathcal{Q}_L=\int d\xi^1 \partial_0 ff^{-1} ,\quad \mathcal{Q}_R=\int d\xi^1 f^{-1}\partial_0 f .
 \label{eq:dressing_review_charges_left_right}
 \end{equation}

\subsection{The Dressing Method}
Let $F=\mathrm{SL}(n,\mathbb C)$ and suppose that we already know a solution $f$ --- the seed solution --- of the equation of motion \eqref{eq:dressing_review_zero_curvature}. The dressing transformation allows us to construct a new solution $f'$ from the seed solution $f$. In principle, we can solve the auxiliary system \eqref{eq:dressing_review_auxiliary} with the condition $\Psi(0)=f$ and find $\Psi(\lambda)$. The dressing transformation involves constructing a new solution $\Psi'(\lambda)$ of the auxiliary system \eqref{eq:dressing_review_auxiliary} of the form
\begin{equation}
\Psi'(\lambda)=\chi(\lambda)\Psi(\lambda) .
\label{eq:dressing_review_dressing_transformation}
\end{equation}
The $n\times n$ matrix $\chi(\lambda)$ is called the dressing factor. It can be shown \cite{Harnad:1983we} that the general form of $\chi$ is
 \begin{equation}
\chi(\lambda) = I + \sum_i\frac{Q_i}{\lambda-\lambda_i} , \quad \chi(\lambda)^{-1} = I + \sum_i\frac{R_i}{\lambda-\mu_i} .
\label{eq:dressing_review_dressing_factor_general}
\end{equation}

It turns out that at the level of the $F=\mathrm{SL}(n,\mathbb C)$ NLSM, the poles can be selected at arbitrary positions on the complex plane and we are left with the problem of specifying the appropriate residues. There are two conditions that the residues must satisfy, which are adequate for their specification. The first one is the demand that $\chi(\lambda)\chi(\lambda)^{-1} = I$. Taking the residues of this equation at the positions of the poles $\lambda_i$ and $\mu_i$ provides a set of algebraic equations for $Q_i$ and $R_i$. Notice that one has to be careful when a pole of $\chi(\lambda)$ coincides with a pole of $\chi(\lambda)^{-1}$, since in this case the product $\chi(\lambda)\chi(\lambda)^{-1} $ will have a second order pole, which has to be considered separately.

The solution $\Psi'\left( \lambda \right)$ of the auxiliary system gives rise to a new solution $f' = \Psi' \left( 0 \right)$ of the NLSM. It follows that $f'$ and $\Psi'\left( \lambda \right)$ must satisfy equations \eqref{eq:dressing_review_auxiliary}, namely,
\begin{equation}
{J'}_\pm=(1\pm\lambda)\partial_\pm \Psi' \left( \lambda \right) \left( {\Psi'} \left( \lambda \right) \right)^{-1}.
\end{equation}
Using \eqref{eq:dressing_review_dressing_transformation} this reduces to
\begin{equation}
{J'}_\pm=(1\pm\lambda)\partial_\pm\chi\chi^{-1}+\chi J_{\pm}\chi^{-1}=-(1\pm\lambda)\chi\partial_\pm\chi^{-1}+\chi J_{\pm}\chi^{-1} .
\label{eq:dressing_review_current_dressed}
\end{equation}
Taking the residues at $\lambda_i$ and $\mu_j$ of the previous equations yields two more relations for the unknown matrices $Q_i$ and $R_i$, being first order differential equations for the latter. These, combined with the algebraic equations derived from the residues of the equation $\chi(\lambda)\chi(\lambda)^{-1} = I$, are sufficient for the specification of the residues $Q_i$ and $R_i$. More details are provided in appendix \ref{sec:dressing_formalities} and in \cite{Harnad:1983we}.

We now turn to the effect of the dressing transformation on the sigma model charge. The latter gets altered by
\begin{equation}
\Delta \mathcal{Q}_L := \int d\xi^1\left({J'}_0-J_0\right)=\frac{1}{2} \int d\xi^1\left({J'}_+ - {J'}_- - J_+ + J_- \right) .
\end{equation}
We notice that the left hand side of \eqref{eq:dressing_review_current_dressed} is independent of $\lambda$. In the limit $|\lambda|\to\infty$ \eqref{eq:dressing_review_current_dressed} reduces to
\begin{equation}
{J'}_\pm = \pm \partial_\pm\sum_jQ_j+ J_{\pm}
\label{eq:dressing_review_current_charge}
\end{equation}
Using \eqref{eq:dressing_review_current_charge} we arrive at the equation
\begin{equation}
\Delta \mathcal{Q}_L = \sum_j \int d \xi^1 \partial_1 Q_j ,
\label{eq:dressing_review_change_of_charge}
\end{equation}
which relates the charges of the seed and dressed solutions.

\subsection{Involutions}
\label{subsec:dressing_involutions}

As it was stated earlier, the previous results refer to the $\mathrm{SL}(n,\mathbb R)$ NLSM. For our purposes $f$ must take values in some symmetric space $F/G$, where $F$, $G$ are Lie groups and $G\subset F$. This can be achieved by constraining appropriately the field $f$ to take values in the coset $F/G$ with the help of an involution. An involution is a bijective mapping $\sigma: \,F\to F$ with the properties
\begin{equation}
\sigma^2=1 , 
\end{equation}
and
\begin{equation}
\quad\sigma(f_1f_2)=\sigma(f_1)\sigma(f_2),\\
\end{equation}
where $f_1,f_2\in F$. Furthermore, we demand that the involution $\sigma$ obeys 
\begin{equation}
\sigma(g)=g, \quad \forall g\in G .
\end{equation}
On the Lie algebra level the mapping $\sigma$ is just a linear operator acting on the vector space $\mathbf f$, having the property $\sigma^2=1$. Since $\sigma^2=1$, $\sigma$ has eigenvalues $\pm1$ and thus the vector space $\mathbf f$ can be decomposed as follows 
\begin{equation}
\mathbf f=\mathbf g\oplus \mathbf p ,
\end{equation}
where $\mathbf g$ and $\mathbf p$ are the $+1$ and $-1$ eigenspaces respectively. Trivially it holds that
\begin{equation}
[\mathbf g,\mathbf g]\subset \mathbf g ,\quad [\mathbf g,\mathbf p]\subset \mathbf p ,\quad [\mathbf p,\mathbf p]\subset\mathbf g ,
\end{equation}
where $\mathbf g$ is by definition the Lie algebra corresponding to the subgroup $G$ and $\mathbf p$ is its orthogonal complement. Thus, the involution $\sigma$ naturally splits the group $F$ to the subgroup $G$ and the coset $F/G$.

We consider now the following coset valued field
\begin{equation}
\mathcal{F}:=\sigma(f)f^{-1} .
\end{equation}
It can be easily shown that it is indeed invariant under the coset equivalence relation $f\sim fg$. Acting on $\mathcal F$ with $\sigma$ gives the following relation
\begin{equation}
\sigma(\mathcal F)=\mathcal F^{-1} .
\label{eq:dress_review_f_in_coset_condition}
\end{equation}
This is the constraint we need to impose on the fields $f$ of the NLSM \eqref{eq:dressing_review_nlsm_action} in order to restrict them inside the coset $F/G$. In the following, we assume that the sigma model field is appropriately constrained into the coset $F/G$ and we denote it again as $f$. The NLSM action with target space the coset $F/G$ is not invariant under the full $F_L\times F_R$ symmetry group, but only under transformations of the type
\begin{equation}
f\to\sigma(U)\,f\,U^{-1} .
\end{equation}
This implies that the conserved charges $\mathcal{Q}_L,\mathcal{Q}_R$ are not independent anymore. They are related by
\begin{equation}
\mathcal{Q}_L=-\sigma(\mathcal{Q}_R) .
\end{equation}

In general, when we want to study the NLSM with a symmetric target space $F/G$, we start with the model on the group $\mathrm{SL}(n,\mathbb C)$. Using one or possibly more involutions denoted by $\sigma_+$, we restrict to the subgroup $F\subset \mathrm{SL}(n,\mathbb C)$ and then via another involution $\sigma_-$ we further restrict the target space to be $F/G\subset F$. In this work we are interested in the spheres $S^n=\mathrm{SO}(n+1)/\mathrm{SO}(n)$. For this purpose, we need three involutions \cite{Pohl_avatars}.

Firstly, we demand invariance ($\sigma_+(f) = f$), under the involution
\begin{equation}
\sigma_+(f)=\left(f^\dagger\right)^{-1} .
\end{equation}
Obviously, this involution restricts the target space to be $\mathrm{SU}(n+1)\subset \mathrm{SL}(n+1,\mathbb C)$. The auxiliary system equations \eqref{eq:dressing_review_auxiliary} and invariance of the group element $f$ under this involution imply that $\Psi(\lambda)$ obeys 
\begin{equation}
\Psi(\lambda)=\left(\Psi(\bar{\lambda})^\dagger\right)^{-1}.
\label{eq:dress_review_condition_unitarity}
\end{equation}
We require that the new solution $f'$, found after the application of the dressing method, also belongs in $\mathrm{SU}(n+1)$. This means that the condition \eqref{eq:dress_review_condition_unitarity} should be obeyed by $\Psi '(\lambda)$, which in turn implies that $\chi(\lambda)=\left(\chi(\bar{\lambda})^\dagger\right)^{-1}$. Applying the above to the dressing factor, as given by equation \eqref{eq:dressing_review_dressing_factor_general}, implies that the poles and the residues obey
\begin{equation}
\mu_i = \bar{\lambda}_i \quad \text{and} \quad R_i = \bar{Q}_i ,
\end{equation}
simplifying the dressing factor $\chi$. The simplest case to consider is a dressing factor with only one pole. In this case, if the initial solution $f$ was the vacuum solution, the dressed one $f'$ turns out to be the one soliton solution. By adding more poles to the dressing factor one would get the $N$-soliton solution in general.

The second involution needed is the following
\begin{equation}
\sigma_-(f)=\theta f\theta^{-1} ,\quad \theta=\text{diag}\{+1,\cdots,+1,-1\} .
\end{equation}
Demanding that $\sigma_-(f)=f^{-1}$, --- see equation \eqref{eq:dress_review_f_in_coset_condition} --- restricts the target space to be $\mathrm{SU}(n+1)/\mathrm{U}(n)$. Then, the auxiliary system \eqref{eq:dressing_review_auxiliary} implies that when $f$ obeys $\sigma_-(f)=f^{-1}$, $\Psi(\lambda )$ obeys
\begin{equation}
\Psi(\lambda^{-1})=f\theta\Psi(\lambda)\theta^{-1} .
\label{eq:dress_review_condition_coset}
\end{equation}
Applying the above on $\Psi'(\lambda )$, results in $\chi(\lambda^{-1})=f'\theta\chi(\lambda)\theta^{-1}f^{-1}$. This in turn implies that poles in the dressing factor come in pairs $\{\lambda, \lambda^{-1}\}$. Thus, the simplest case to consider is that of a dressing factor with two poles $\lambda_1$ and $\lambda_2 = 1 / \lambda_1$. In this case, the corresponding residues must satisfy 
\begin{equation}
Q_2 = - \lambda_2^2 f' \theta Q_1 \theta f .
\end{equation}

Finally, we demand invariance of $f$ under the involution
\begin{equation}
\sigma_+(f)=f^* .
\end{equation}
This is just the reality condition to be imposed on the solution, so that it belongs to the coset $\mathrm{SO}(n+1)/\mathrm{SO}(n)$. The auxiliary system \eqref{eq:dressing_review_auxiliary} implies that $\Psi ( \lambda )$ must obey
\begin{equation}
\overline{\Psi\left( \bar{\lambda} \right) } = \Psi \left( \lambda \right) .
\end{equation}
Demanding the above for $\Psi ' ( \lambda )$ leads to the fact that the poles in the dressing factor must come in pairs $\{\lambda, \bar{\lambda}\}$. Had we imposed this involution to the $\mathrm{SU}(N)$ model, we would have concluded that the simplest possible dressing factor would have two poles $\lambda_1$ and $\lambda_2 = \bar{\lambda}_1$ with the corresponding residues obeying
\begin{equation}
Q_2 = \bar{Q}_1 .
\end{equation}

Notice that imposing the reality involution together with the unitarity involution adds an extra complexity to finding the appropriate dressing factor. The latter involution enforces the poles of $\chi \left( \lambda \right)$ to come in pairs of numbers being complex conjugate to each other. The former involution enforces the poles of $\chi \left( \lambda \right) ^{-1}$ to be the complex conjugates of the poles of $\chi \left( \lambda \right)$. Thus, when studying $\mathrm{SO}(N)$ models or coset subspaces of the latter, the dressing factor $\chi \left( \lambda \right)$ necessarily has poles that coincide with the poles of its inverse $\chi \left( \lambda \right) ^{-1}$, complicating the specification of the residues $Q_i$ as we discussed above. In the simplest case of two poles, it obviously holds that $\mu_1 = \bar{\lambda}_1 = \lambda_2$ and $\mu_2 = \bar{\lambda}_2 = \lambda_1$.

In the case of interest, we have to impose the constraints originating from the coset involution $\sigma_-$ and the reality involution. This implies that naively, the dressing factor in the case of the $\mathrm{SO}(n+1)/\mathrm{SO}(n)$ NLSM comes with quadruplets of poles $\{\lambda_1, \lambda_2 = \bar{\lambda}, \lambda_3 = \lambda^{-1}, \lambda_4 = \bar{\lambda}^{-1}\}$, with residues obeying $Q_2 = \bar{Q}_1$, $Q_3 = - \lambda_2^2 f' \theta Q_1 \theta f$ and $Q_4 = \bar{Q}_3$. However, the simplest possible dressing factor does not have four poles, but only two. When $|\lambda_1| = 1$, it holds that $\bar{\lambda}= \lambda^{-1}$ and the quadruplet reduces to a doublet of poles. This is the case that we will consider from now on. In this case, the dressing factor assumes the form
\begin{equation}
\chi \left( \lambda  \right) = I + \frac{{{\lambda _1} - {{\bar \lambda }_1}}}{{\lambda  - {\lambda _1}}}P + \frac{{{{\bar \lambda }_1} - {\lambda _1}}}{{\lambda  - {{\bar \lambda }_1}}}\bar P ,
\label{eq:dress_review_dressing_factor_two_conjugate_poles}
\end{equation}
where
\begin{equation}
P = \frac{{\Psi \left( {{{\bar \lambda }_1}} \right)p{p^\dag }{\Psi ^{ - 1}}\left( {{\lambda _1}} \right)}}{{{p^\dag }{\Psi ^{ - 1}}\left( {{\lambda _1}} \right)\Psi \left( {{{\bar \lambda }_1}} \right)p}}
\label{eq:dress_review_dressing_factor_two_conjugate_poles_projector}
\end{equation}
and the vector $p$ is any constant complex vector obeying ${p^T}p = 0$ and $\bar p = \theta p$. More details are provided in the appendix \ref{sec:dressing_formalities} and in \cite{Harnad:1983we,SaintAubin:1982wrd}.

\subsection{The Mapping from Unit Vectors to Orthogonal Matrices}
\label{subsec:mapping_X_f}

We map any vector $X$ on the unit sphere S$^n$ to an element $f$ of the $\mathrm{SO}(n+1)/\mathrm{SO}(n)$ coset, via \cite{Pohl_avatars}
\begin{equation}
f = \left( I - 2{X_0}X_0^T \right) \left( {I - 2X{X^T}} \right) ,
\label{eq:dressing_mapping_coset}
\end{equation}
where $X_0$ is a given constant vector with unit norm. This trivially transforms the NLSM action \eqref{eq:dressing_review_nlsm_action} to the string action \eqref{eq:Pohlmeyer_action}. In the following, we denote
\begin{equation}
\theta := I - 2{X_0}X_0^T 
\end{equation}
and in our S$^2$ applications, we will select
\begin{equation}
{X_0} = \left( {\begin{array}{*{20}{c}}
0\\
0\\
1
\end{array}} \right),\quad \theta  = \diag\left\{ { + 1, +1, -1} \right\} ,
\end{equation}
unless otherwise specified. For any unit vector $X$, it is true that
\begin{equation}
\left( {I - 2X{X^T}} \right)\left( {I - 2X{X^T}} \right) = I ,
\end{equation}
implying that ${\theta ^2} = I$. Additionally, since ${f^T} = f$, the above implies that $f$ is an orthogonal matrix obeying ${f^T} = {f^{ - 1}}$. Moreover, notice that $\det \left( {I - 2X{X^T}} \right) =  - 1$, implying that $\det f = 1$ and thus $f \in \textrm{SO}(n+1)$. Finally, $\sigma_- \left( f \right) := \theta f \theta^{-1} = f^{-1}$, implying that $f\in \textrm{SO}(n+1) / \textrm{SO}(n)$.

Let $\alpha$ be the angle between the unit vectors $X_0$ and $X$. Then, the special orthogonal matrix $f$ represents a rotation in the plane defined by $X_0$ and $X$ by an angle equal to $2\alpha$. The matrix $f$ has one real eigenvector $\chi_0 = X_0 \times X$ with eigenvalue equal to one and two complex eigenvectors $\chi_\pm = - e^{\pm i \alpha} X_0 + X$, obeying $\chi_\pm ^T \chi_\pm = 0$, with eigenvalues $e^{\pm 2 i \alpha}$, respectively.

\subsection{Pohlmeyer Reduction and Virasoro Constraints}
\label{subsec:dress_review_Virasoro}

As it was described in \cite{Grigoriev:2007bu} the sigma model on a symmetric space admits a Pohlmeyer reduction, which amounts to exploiting the conformal symmetry of the NLSM at the classical level in order to set the components of the energy momentum tensor to be constant, i.e.
\begin{equation}
T_{\pm\pm}=m_\pm^2 .
\label{eq:dressing_review_Virasoro}
\end{equation}

It was shown in \cite{Miramontes:2008wt} that at an algebraic level the Pohlmeyer reduction is equivalent to imposing the following condition,
\begin{equation}
\partial_\pm f f^{-1}=\xi_\pm\Lambda_\pm\xi_\pm^{-1} ,\quad\text{with}\quad\sigma(\xi_\pm)=f^{-1}\xi_{\pm} ,
\label{eq:dressing_review_Pohlmeyer}
\end{equation} 
where $\Lambda_\pm$ are constant elements in a maximal abelian subspace of  $\mathbf{p}$ and $\xi_\pm\in F$. The degree of freedom left after the reduction is $\gamma= \xi_-^{-1}\xi_+$. In order to see how this is equivalent to \eqref{eq:dressing_review_Virasoro}, we will use the parametrization \eqref{eq:dressing_mapping_coset} for the coset element $f$. The components of the energy momentum tensor of the NLSM are
\begin{equation}
T_{\pm\pm}=\tr(J_\pm J_\pm) .
\end{equation}
From \eqref{eq:dressing_review_currents_left_right}, \eqref{eq:dressing_review_Pohlmeyer} and \eqref{eq:dressing_mapping_coset}, it follows that
\begin{equation}
T_{\pm\pm}=-8(\partial_\pm X^m)(\partial_\pm X^m)= \tr\Lambda_\pm^2 .
\label{eq:dressing_review_Virasoro_final}
\end{equation}
If we make the identification $\tr \Lambda_\pm^2=-8m_\pm^2$, equation \eqref{eq:dressing_review_Virasoro_final} will become \eqref{eq:dressing_review_Virasoro}. This indicates the equivalence between \eqref{eq:dressing_review_Pohlmeyer} and \eqref{eq:dressing_review_Virasoro}. More details on this can be found in \cite{Miramontes:2008wt}.

In order to see if the dressing transformation is compatible with Pohlmeyer reduction, we go back to \eqref{eq:dressing_review_current_dressed}, divide by $(1\pm\lambda)$ and find the residues at $\lambda=\pm1$. This gives the following relations
\begin{equation}
\partial_\pm\tilde f \tilde f^{-1}=\chi(\mp1)\partial_\pm f f^{-1}\chi(\mp1)^{-1} .
\end{equation}
Using equation \eqref{eq:dressing_review_Pohlmeyer} yields
\begin{equation}
\partial_\pm\tilde f \tilde f^{-1}=\chi(\mp1)\xi_\pm \Lambda_\pm\xi_\pm^{-1}\chi(\mp1)^{-1} .
\end{equation}
Therefore, if we set 
\begin{equation}
\tilde\xi_\pm=\chi(\mp1)\xi_\pm\Xi , \quad [\Xi,\Lambda_\pm]=0 ,
\label{eq:dressing_review_Virasoro_xis}
\end{equation}
equation \eqref{eq:dressing_review_Virasoro_xis} will take the form of the Pohlmeyer constraint \eqref{eq:dressing_review_Pohlmeyer}. This shows that the dressing procedure respects the constraint \eqref{eq:dressing_review_Pohlmeyer} or equivalently \eqref{eq:dressing_review_Virasoro}. The element $\Xi$ will be chosen so that the degree of freedom of the reduced system $\tilde\gamma=\tilde\xi_-^{-1}\tilde\xi_+$ is an element of the subgroup $G$.

Interpreting $X^i$ as the coordinates of a string moving on a sphere, it can be shown that the NLSM charge is related to the angular momentum of the string. Using \eqref{eq:dressing_review_Virasoro_final} and \eqref{eq:dressing_review_charges_left_right} we find that 
\begin{equation}
\mathcal{Q}_L = - 2\int\,d\xi^1\left(X^\mu\partial_0 X^\nu-X^\nu\partial_0X^\mu\right) .
\label{eq:review_J_as_NLSM_charge}
\end{equation}
Therefore, the sigma model charge is proportional to the string angular momentum.

\setcounter{equation}{0}
\section{Dressed Elliptic String Solutions}
In this section, we apply the dressing method that we reviewed in section \ref{sec:dress_review}, to the elliptic string solutions of section \ref{sec:review}, using the simplest possible dressing factor, in order to construct new classical string solutions propagating on $\mathbb{R}^t \times$S$^2$.

The non-trivial seed solution of section \ref{sec:review} (equation \eqref{eq:elliptic_solutions_review}) renders the straightforward application of the dressing method very difficult. This is due to the corresponding auxiliary system, which is a complicated system of coupled partial differential equations with non-constant coefficients. In order to avoid these difficulties, we implement an intuitive detour, by expressing the seed solution as a worldsheet dependent rotation matrix, acting on a constant vector, which coincides with the rotation axis of the seed solution, i.e. the $z$-axis. Furthermore, the parametrization of the coset $\mathrm{SO}(3)/\mathrm{SO}(2)$ is carried out, so that this constant vector corresponds to its identity element via the mapping \eqref{eq:dressing_mapping_coset}. In this way, we manage to express one of the two PDEs of the auxiliary system in a form where one of the two worldsheet coordinates does not appear explicitly, making the solution of the system possible. Simultaneously, all components of the auxiliary field equations acquire a given parity under the inversion $\lambda \to 1 / \lambda$, facilitating the application of the coset involution. Finally, the expression of the seed solution as a rotation matrix acting on a constant vector simplifies the translation of the dressed solution from the form of a coset element to a unit vector.

\label{sec:dressed_strings}
\subsection{The Auxiliary System for an Elliptic Seed Solution}

In order to implement the dressing method, we have to solve the auxiliary system \eqref{eq:dressing_review_auxiliary}. This reads
\begin{equation}
{\partial _ \pm }\Psi \left( \lambda  \right) = \frac{1}{{1 \pm \lambda }}\left( {{\partial _ \pm }f} \right){f^{ - 1}}\Psi \left( \lambda  \right) ,
\label{eq:dress_ell_auxiliary}
\end{equation}
where $f$ is a given seed solution of the NLSM and $\Psi \left( \lambda  \right)$ must obey the condition $\Psi \left( 0  \right) = f$. As seed solutions, we are going to use the $\mathrm{SO}(3)/\mathrm{SO}(2)$ coset elements $f$ corresponding to the elliptic string solutions \eqref{eq:elliptic_solutions_review} through the mapping \eqref{eq:dressing_mapping_coset}. These solutions depend in a trivial manner on either the time-like or space-like worldsheet coordinate. It follows that it is technically advantageous to express the auxiliary system \eqref{eq:dress_ell_auxiliary} as a system of differential equations with independent variables the time-like and space-like coordinates $\xi^0$ and $\xi^1$, instead of the left- and right-moving coordinates $\xi^\pm$. Following these lines, the auxiliary system assumes the form
\begin{equation}
{\partial _i}\Psi \left( \lambda  \right) = \left( {{{\tilde \partial }_i}f} \right){f^{ - 1}}\Psi \left( \lambda  \right) ,
\label{eq:dress_ell_auxiliary_01}
\end{equation} 
where $i=0,1$ and
\begin{equation}
{{\tilde \partial }_0} = \frac{1}{{1 - {\lambda ^2}}}{\partial _0} - \frac{\lambda }{{1 - {\lambda ^2}}}{\partial _1},\quad {{\tilde \partial }_1} = \frac{1}{{1 - {\lambda ^2}}}{\partial _1} - \frac{\lambda }{{1 - {\lambda ^2}}}{\partial _0} .
\end{equation}

It turns out to be convenient to express the initial solution $X$ as an orthogonal matrix $U \left( \xi^0 , \xi^1 \right)$ acting on another unit vector $\hat X$, as
\begin{equation}
X := U\hat X .
\end{equation}
It has to be noted that $\hat X$ is not a solution of the NLSM. In terms of the vector $\hat X$, the seed solution $f$ reads
\begin{equation}
f = \theta U\theta \hat f{U^T} ,
\end{equation}
where
\begin{equation}
\hat f: = \theta \left( {I - 2\hat X{{\hat X}^T}} \right) .
\end{equation}
Obviously $\hat f \in \textrm{SO}(3)$. It is also convenient to define $\hat \Psi \left( \lambda  \right)$ as
\begin{equation}
\Psi \left( \lambda  \right): = \theta U\theta \hat \Psi \left( \lambda  \right) .
\end{equation}
Then, the equations of the auxiliary system \eqref{eq:dress_ell_auxiliary_01}, expressed in terms of hatted quantities, assume the form
\begin{equation}
{\partial _i}\hat \Psi  = \left[ {\theta {U^T}\left( {\left( {{{\tilde \partial }_i} - {\partial _i}} \right)U} \right)\theta  - \hat f{U^T}\left( {{{\tilde \partial }_i}U} \right){{\hat f}^T} + \left( {{{\tilde \partial }_i}\hat f} \right){{\hat f}^T}} \right]\hat \Psi .
\end{equation}

One can always select the orthogonal matrix $U$ so that $\hat X = {X_0}$. For this specific selection, $\hat f = I$ and the equations of the auxiliary system get simplified to the form
\begin{equation}
{\partial _i}\hat \Psi  = \left[ {\theta {U^T}\left( {\left( {{{\tilde \partial }_i} - {\partial _i}} \right)U} \right)\theta  - {U^T}\left( {{{\tilde \partial }_i}U} \right)} \right]\hat \Psi .
\label{eq:dress_auxiliary_ell_hatted}
\end{equation}
Furthermore, the condition $\Psi \left( 0 \right) = f$ translates to the condition $\hat \Psi \left( 0 \right) = U^T$.

Without loss of generality, we perform the analysis in the case of seed solutions with static Pohlmeyer counterparts. The latter read
\begin{equation}
X = \left( {\begin{array}{*{20}{c}}
{{F_1}\left( {{\xi^1}} \right)\cos \varphi \left( {{\xi^0},{\xi^1}} \right)}\\
{{F_1}\left( {{\xi^1}} \right)\sin \varphi \left( {{\xi^0},{\xi^1}} \right)}\\
{{F_2}\left( {{\xi^1}} \right)}
\end{array}} \right) ,
\label{eq:dress_X_elliptic_F12}
\end{equation}
where
\begin{align}
{F_1}\left( {{\xi^1}} \right) &= \sqrt {\frac{{\wp \left( {{\xi^1} + {\omega _2}} \right) - \wp \left( a \right)}}{{{x_1} - \wp \left( a \right)}}} , \label{eq:dressed_strings_F1}\\
{F_2}\left( {{\xi^1}} \right) &= \sqrt {\frac{{{x_1} - \wp \left( {{\xi^1} + {\omega _2}} \right)}}{{{x_1} - \wp \left( a \right)}}} , \label{eq:dressed_strings_F2}\\
\varphi \left( {{\xi^0},{\xi^1}} \right) &= \sqrt {{x_1} - \wp \left( a \right)} {\xi^0} - \Phi \left( \xi^1 ; a \right) .
\end{align}
Notice that $F_1$ and $F_2$ obey $F_1^2\left( {{\xi^1}} \right) + F_2^2\left( {{\xi^1}} \right) = 1$. Moreover, $F_1$, $F_2$ and $\varphi$ satisfy
\begin{align}
{\partial _0}\varphi  = \sqrt {{x_1} - \wp \left( a \right)} ,& \quad {\partial _1}\varphi  =  - \frac{{i\wp '\left( a \right)}}{2}\frac{1}{{\wp \left( {{\xi^1} + {\omega _2}} \right) - \wp \left( a \right)}} ,\\
{\partial _0}{F_1} = 0,& \quad {\partial _0}{F_2} = 0 ,\\
{\partial _1}{F_1} = \frac{{{F_3}}}{{{F_1}}} ,& \quad {\partial _1}{F_2} =  - \frac{{{F_3}}}{{{F_2}}} ,
\end{align}
where
\begin{equation}
{F_3}\left( {{\xi^1}} \right) := \frac{{\wp '\left( {{\xi^1} + {\omega _2}} \right)}}{{2\left( {{x_1} - \wp \left( a \right)} \right)}} .
\end{equation}

In terms of the functions $F_1$, $F_2$ and $\varphi$, the Virasoro constraints are expressed as
\begin{align}
F_1^2\left[ {{{\left( {{\partial _0}\varphi } \right)}^2} + {{\left( {{\partial _1}\varphi } \right)}^2}} \right] + {\left[ {{F_2}\left( {{\partial _1}{F_1}} \right) - {F_1}\left( {{\partial _1}{F_2}} \right)} \right]^2} &= \frac{{m_ + ^2 + m_ - ^2}}{2} , \label{eq:ffphi_Virasoro_1}\\
2F_1^2\left( {{\partial _0}\varphi } \right)\left( {{\partial _1}\varphi } \right) &= \frac{{m_ + ^2 - m_ - ^2}}{2} .\label{eq:ffphi_Virasoro_2}
\end{align}
Similarly, the equations of motion imply
\begin{align}
{F_1}\partial _1^2\varphi  + 2\left( {{\partial _1}{F_1}} \right)\left( {{\partial _1}\varphi } \right) &= 0 , \label{ffphi_eom_1}\\
{F_2}\partial _1^2{F_1} - {F_1}\partial _1^2{F_2} &= {F_1}{F_2}\left[ { - {{\left( {{\partial _0}\varphi } \right)}^2} + {{\left( {{\partial _1}\varphi } \right)}^2}} \right] , \label{ffphi_eom_2}\\
{F_1}\partial _1^2{F_1} + {F_2}\partial _1^2{F_2} &=  - {\left[ {{F_2}\left( {{\partial _1}{F_1}} \right) - {F_1}\left( {{\partial _1}{F_2}} \right)} \right]^2} . \label{ffphi_eom_3}
\end{align}

Equation \eqref{eq:dress_X_elliptic_F12} implies that the seed elliptic string solution can be expressed as $X = U{X_0}$, where
\begin{equation}
U = {U_2}{U_1} 
\end{equation}
and the matrices $U_1$ and $U_2$ are given by
\begin{equation}
{U_1} = \left( {\begin{array}{*{20}{c}}
{{F_2}}&0&{{F_1}}\\
0&1&0\\
{{-F_1}}&0&{{F_2}}
\end{array}} \right),\quad {U_2} = \left( {\begin{array}{*{20}{c}}
{\cos \varphi }&{ - \sin \varphi }&0\\
{\sin \varphi }&{\cos \varphi }&0\\
0&0&1
\end{array}} \right) .
\end{equation}
The equations of the auxiliary system require the calculation of the quantities
\begin{equation}
{U^T}\left( {\partial_i U} \right) = U_1^TU_2^T\left( {\partial_i {U_2}} \right){U_1} + U_1^T\left( {\partial_i {U_1}} \right) .
\end{equation}
It is a matter of simple algebra to show that
\begin{eqnarray}
U_1^TU_2^T\left( {{\partial _i}{U_2}} \right){U_1} = \left( {{\partial _i}\varphi } \right)U_1^T{T_3}{U_1} = \left( {{\partial _i}\varphi } \right)\left( {{F_2}{T_3} + {F_1}{T_1}} \right) , \label{eq:dress_ell_U_der1} \\
U_1^T\left( {{\partial _0}{U_1}} \right) = O , \quad U_1^T\left( {{\partial _1}{U_1}} \right) = \left[ {F_2}\left( {{\partial _1}{F_1}} \right) - {F_1}\left( {{\partial _1}{F_2}} \right) \right] {T_2} = \frac{{{F_3}}}{{{F_1}{F_2}}}{T_2} , \label{eq:dress_ell_U_der2}
\end{eqnarray}
where $T_i$ are the $\mathrm{SO}(3)$ generators, namely,
\begin{equation}
{T_1} = \left( {\begin{array}{*{20}{c}}
0&0&0\\
0&0&{ - 1}\\
0&1&0
\end{array}} \right),\quad {T_2} = \left( {\begin{array}{*{20}{c}}
0&0&1\\
0&0&0\\
{ - 1}&0&0
\end{array}} \right),\quad {T_3} = \left( {\begin{array}{*{20}{c}}
0&{ - 1}&0\\
1&0&0\\
0&0&0
\end{array}} \right) .
\end{equation}
Adopting the notation
\begin{equation}
{U^T}\left( {{\partial _i}U} \right) = k_i^j{T_j} ,
\end{equation}
the equations \eqref{eq:dress_ell_U_der1} and \eqref{eq:dress_ell_U_der2} imply that
\begin{align}
k_0^1 = - \left( {{\partial _0}\varphi } \right){F_1},&\quad k_1^1 = - \left( {{\partial _1}\varphi } \right){F_1} , \label{eq:defs_k_1}\\
k_0^2 = 0,&\quad k_1^2 = {F_2}\left( {{\partial _1}{F_1}} \right) - {F_1}\left( {{\partial _1}{F_2}} \right) = \frac{{{F_3}}}{{{F_1}{F_2}}} ,\label{eq:defs_k_2}\\
k_0^3 = \left( {{\partial _0}\varphi } \right){F_2},&\quad k_1^3 = \left( {{\partial _1}\varphi } \right){F_2} .\label{eq:defs_k_3}
\end{align}
Notice that none of the coefficients $k_i^j$ depends on the time-like coordinate $\xi^0$.

Similarly, we adopt the notation
\begin{equation}
{\partial _i}\hat \Psi  = \kappa _i^j{T_j}\hat \Psi .
\label{eq:dress_auxiliary_ell_cross_product}
\end{equation}
Observing that
\begin{equation}
\theta {T_1}\theta  = - {T_1},\quad \theta {T_2}\theta  =  - {T_2},\quad \theta {T_3}\theta  = {T_3} ,
\end{equation}
the equations of the auxiliary system \eqref{eq:dress_auxiliary_ell_hatted} imply that
\begin{align}
\kappa _{0/1}^3 &= - k_{0/1}^3 ,\label{eq:defs_kappa_1}\\
\kappa _{0/1}^{1/2} &=  - \frac{{1 + {\lambda ^2}}}{{1 - {\lambda ^2}}}k_{0/1}^{1/2} + \frac{{2\lambda }}{{1 - {\lambda ^2}}}k_{1/0}^{1/2} \nonumber \\
&=  - \coth z k_{0/1}^{1/2} + \csch z k_{1/0}^{1/2} ,\label{eq:defs_kappa_2}
\end{align}
where $\lambda = e^z$. The above imply that the coefficients $\kappa_i^j$ obey the properties
\begin{align}
\kappa _{0/1}^3\left( {1/\lambda } \right) &= \kappa _{0/1}^3\left( \lambda  \right) ,\\
\kappa _{0/1}^{1/2}\left( {1/\lambda } \right) &= - \kappa _{0/1}^{1/2}\left( \lambda  \right) 
\end{align}
or in a shorthand notation
\begin{equation}
\kappa_{0/1} \left( 1 / \lambda \right) = - \theta \kappa_{0/1} \left( \lambda \right),
\end{equation}
where
\begin{equation}
{\kappa _{0/1}} = \left( {\begin{array}{*{20}{c}}
{\kappa _{0/1}^1}\\
{\kappa _{0/1}^2}\\
{\kappa _{0/1}^3}
\end{array}} \right) .
\end{equation}

It is a matter of algebra to show that $\kappa _0^T{\kappa _0}$ equals
\begin{multline}
\kappa _0^T{\kappa _0} := \Delta = {\left( {{\partial _0}\varphi } \right)^2} - 2F_1^2\left( {{\partial _0}\varphi } \right)\left( {{\partial _1}\varphi } \right)\frac{{1 + {\lambda ^2}}}{{1 - {\lambda ^2}}}\frac{{2\lambda }}{{1 - {\lambda ^2}}}\\
 + \left\{ {F_1^2\left[ {{{\left( {{\partial _0}\varphi } \right)}^2} + {{\left( {{\partial _1}\varphi } \right)}^2}} \right] + {{\left[ {{F_2}\left( {{\partial _1}{F_1}} \right) - {F_1}\left( {{\partial _1}{F_2}} \right)} \right]}^2}} \right\}{\left( {\frac{{2\lambda }}{{1 - {\lambda ^2}}}} \right)^2}.
\end{multline}
Using the Virasoro constraints \eqref{eq:ffphi_Virasoro_1} and \eqref{eq:ffphi_Virasoro_2}, we can express $\Delta$ in terms of the quantities $E$ and $m_\pm$,
\begin{equation}
\begin{split}
\Delta &= \frac{E}{2} + \frac{m_ + ^2}{4}{\left( {\frac{{1 - \lambda }}{{1 + \lambda }}} \right)^2} + \frac{m_ - ^2}{4}{\left( {\frac{{1 + \lambda }}{{1 - \lambda }}} \right)^2} \\
 &= \frac{E}{2} + \frac{{m_ + ^2}}{4}{\tanh ^2}\frac{z }{2} + \frac{{m_ - ^2}}{4}{\coth ^2}\frac{z }{2}.
\end{split}
\end{equation} 
Thus, the quantity $\Delta$ is a constant. Notice also that $\Delta\left( {1/\lambda } \right) = \Delta\left( \lambda  \right)$. The quantity $\Delta$ could be considered as the generalization of the parameter $\ell^2$ of the elliptic seed solution after a ``boost'' in the worldsheet coordinates with complex rapidity $z / 2$.

\subsection{The Solution of the Auxiliary System}

Since all coefficients in the equations of the auxiliary system \eqref{eq:dress_auxiliary_ell_cross_product} are functions of $\xi^1$ only, we may proceed to solve those that involve the derivatives of $\hat \Psi$ with respect to $\xi^0$ as ordinary differential equations, upgrading the undetermined constants to undetermined functions of $\xi^1$. These equations are a set of three identical linear first order systems, one for each column of $\hat \Psi$, $\hat \Psi_i$, $i=1,2,3$. This linear system has the solution
\begin{equation}
\hat \Psi_i \left( \lambda  \right) = {c_i^0}\left( {{\xi^1}} \right){v_0} + {c_i^ + }\left( {{\xi^1}} \right){v_ + }{e^{i\sqrt{\Delta} {\xi^0}}} + {c_i^ - }\left( {{\xi^1}} \right){v_ - }{e^{ - i\sqrt{\Delta} {\xi^0}}} ,
\end{equation}
where
\begin{equation}
{v_0} = \frac{1}{{\sqrt{\Delta} }}\left( {\begin{array}{*{20}{c}}
{\kappa _0^1}\\
{\kappa _0^2}\\
{\kappa _0^3}
\end{array}} \right),\quad {v_ \pm } = \frac{1}{{\sqrt {\Delta \left( {{{\left( {\kappa _0^1} \right)}^2} + {{\left( {\kappa _0^2} \right)}^2}} \right)} }}\left( {\begin{array}{*{20}{c}}
{\kappa _0^3\kappa _0^1 \pm i\sqrt{\Delta} \kappa _0^2}\\
{\kappa _0^3\kappa _0^2 \mp i\sqrt{\Delta} \kappa _0^1}\\
{ - {{\left( {\kappa _0^1} \right)}^2} - {{\left( {\kappa _0^2} \right)}^2}}
\end{array}} \right) .
\end{equation}
The vectors $v_0$ and $v_\pm$ have been selected so that $v_0^T {v_0} = 1$, whereas $v_\pm^T{ v_\pm} = 0$. Furthermore $v_\pm$ obey $\left( \frac{ v_+ + v_-}{2} \right)^T \left(\frac{ v_+ + v_-}{2} \right)  = \left( \frac{ v_+ - v_-}{2i} \right)^T \left( \frac{ v_+ - v_-}{2i} \right)= 1$.

Using the definitions \eqref{eq:defs_k_1}, \eqref{eq:defs_k_2} and \eqref{eq:defs_k_3}, as well as the equations of motion \eqref{ffphi_eom_1}, \eqref{ffphi_eom_2} and \eqref{ffphi_eom_3}, it is a matter of algebra to show that
\begin{align}
{\partial _1}k_0^1 =  - k_1^2k_0^3,&\quad {\partial _1}k_1^1 = k_1^3k_1^2 , \label{eq:dressed_string_k1_der}\\
{\partial _1}k_0^2 = 0,&\quad {\partial _1}k_1^2 =  - k_1^1k_1^3 + k_0^1k_0^3 , \label{eq:dressed_string_k2_der}\\
{\partial _1}k_0^3 = k_1^2k_0^1,&\quad {\partial _1}k_1^3 = k_1^2k_1^1 + 2 k_1^3k_1^2k_0^3/k_0^1 .
\end{align}
Then, the definitions \eqref{eq:defs_kappa_1} and \eqref{eq:defs_kappa_2} imply that
\begin{align}
{\partial _1}\kappa _0^1 &= \kappa _1^2\kappa _0^3 - \kappa _1^3\kappa _0^2 , \\
{\partial _1}\kappa _0^2 &= \kappa _1^3\kappa _0^1 - \kappa _1^1\kappa _0^3 , \\
{\partial _1}\kappa _0^3 &= \kappa _1^1\kappa _0^2 - \kappa _1^2\kappa _0^1
\end{align}
or in a shorthand notation
\begin{equation}
{\partial _1}{{ \kappa }_0} = {{ \kappa }_1} \times {{ \kappa }_0} .
\end{equation}
The vectors $v_0$ and $v_\pm$ can be written in terms of ${{ \kappa }_0}$ as
\begin{align}
{v_0} &= \frac{{{\kappa _0}}}{{\sqrt {\kappa _0^T{\kappa _0}} }}: = {e_3},\\
{v_ \pm } &= \frac{{{X_0} \times {\kappa _0}}}{{\sqrt {{{\left( {{X_0} \times {\kappa _0}} \right)}^T}\left( {{X_0} \times {\kappa _0}} \right)} }} \times \frac{{{\kappa _0}}}{{\sqrt {\kappa _0^T{\kappa _0}} }} \mp i\frac{{{X_0} \times {\kappa _0}}}{{\sqrt {{{\left( {{X_0} \times {\kappa _0}} \right)}^T}\left( {{X_0} \times {\kappa _0}} \right)} }}: = {e_1} \mp i{e_2} .
\end{align}
The vectors
\begin{equation}
{e_i} = \left\{ {\frac{{{X_0} \times {\kappa _0}}}{{\sqrt {{{\left( {{X_0} \times {\kappa _0}} \right)}^T}\left( {{X_0} \times {\kappa _0}} \right)} }} \times \frac{{{\kappa _0}}}{{\sqrt {\kappa _0^T{\kappa _0}} }},\frac{{{X_0} \times {\kappa _0}}}{{\sqrt {{{\left( {{X_0} \times {\kappa _0}} \right)}^T}\left( {{X_0} \times {\kappa _0}} \right)} }},\frac{{{\kappa _0}}}{{\sqrt {\kappa _0^T{\kappa _0}} }}} \right\} 
\end{equation}
form a basis, which obeys $e_i^T {e_j} = \delta_{ij}$ and  $e_i \times e_j = \varepsilon_{ijk}e_k$. Notice that as $\lambda \to 0$,
\begin{equation}
{e_1}\left( 0 \right) = \left( {\begin{array}{*{20}{c}}
{ - {F_2}}\\
0\\
{ - {F_1}}
\end{array}} \right),\quad  {e_2}\left( 0 \right) = \left( {\begin{array}{*{20}{c}}
0\\
1\\
0
\end{array}} \right),\quad {e_3}\left( 0 \right) = \left( {\begin{array}{*{20}{c}}
{{F_1}}\\
0\\
{ - {F_2}}
\end{array}} \right)
\end{equation}
and furthermore
\begin{equation}
{e_{1/2}}\left( {1/\lambda } \right) = \theta {e_{1/2}}\left( {\lambda } \right),\quad {e_3}\left( {1/\lambda } \right) =  - \theta {e_3}\left( {\lambda } \right)  .
\label{eq:dress_ell_e_invertion}
\end{equation}

Using the fact that $\kappa_0^T {\kappa_0}$ is constant, one can show that
\begin{align}
{\partial _1}{e_1} - {\kappa _1} \times {e_1} &=  - \sqrt {\kappa _0^T{\kappa _0}} \frac{{{{\left( {{X_0} \times {\kappa _1}} \right)}^T}\left( {{X_0} \times {\kappa _0}} \right)}}{{{{\left( {{X_0} \times {\kappa _0}} \right)}^T}\left( {{X_0} \times {\kappa _0}} \right)}}{e_2} ,\\
{\partial _1}{e_2} - {\kappa _1} \times {e_2} &= \sqrt {\kappa _0^T{\kappa _0}} \frac{{{{\left( {{X_0} \times {\kappa _1}} \right)}^T}\left( {{X_0} \times {\kappa _0}} \right)}}{{{{\left( {{X_0} \times {\kappa _0}} \right)}^T}\left( {{X_0} \times {\kappa _0}} \right)}}{e_1} ,\\
{\partial _1}{e_3} - {\kappa _1} \times {e_3} &= 0 ,
\end{align}
implying that
\begin{align}
{\partial _1}{v_0} - {\kappa _1} \times {v_0} &= 0,\\
{\partial _1}{v_ \pm } - {\kappa _1} \times {v_ \pm } &=  \mp i\sqrt {\kappa _0^T{\kappa _0}} \frac{{{{\left( {{X_0} \times {\kappa _1}} \right)}^T}\left( {{X_0} \times {\kappa _0}} \right)}}{{{{\left( {{X_0} \times {\kappa _0}} \right)}^T}\left( {{X_0} \times {\kappa _0}} \right)}}{v_ \pm }: =  \mp ig\left( {{\xi ^1}} \right){v_ \pm },
\end{align}
where
\begin{equation}
g\left( {{\xi^1}} \right) = \sqrt{\Delta} \frac{{\kappa _1^1\kappa _0^1 + \kappa _1^2\kappa _0^2}}{{{{\left( {\kappa _0^1} \right)}^2} + {{\left( {\kappa _0^2} \right)}^2}}} .
\end{equation}
It is a matter of algebra to show that
\begin{equation}
g\left( {{\xi^1}} \right) = \frac{{\sqrt{\Delta} \left( {\frac{{m_ + ^2}}{4}{{\left( {\frac{{1 - \lambda }}{{1 + \lambda }}} \right)}^2} - \frac{{m_ - ^2}}{4}{{\left( {\frac{{1 + \lambda }}{{1 - \lambda }}} \right)}^2}} \right)}}{{\wp \left( {{\xi^1} + {\omega _2}} \right) + \frac{E}{6} + \frac{{m_ + ^2}}{4}{{\left( {\frac{{1 - \lambda }}{{1 + \lambda }}} \right)}^2} + \frac{{m_ - ^2}}{4}{{\left( {\frac{{1 + \lambda }}{{1 - \lambda }}} \right)}^2}}} = - \frac{i}{2}\frac{{\wp '\left( \tilde{a} \right)}}{{\wp \left( {{\xi^1} + {\omega _2}} \right) - \wp \left( \tilde{a} \right)}} ,
\end{equation}
where
\begin{equation}
\begin{split}
\wp \left( \tilde{a} \right) &=  - \frac{E}{6} - \frac{{m_ + ^2}}{4}{\left( {\frac{{1 - \lambda }}{{1 + \lambda }}} \right)^2} - \frac{{m_ - ^2}}{4}{\left( {\frac{{1 + \lambda }}{{1 - \lambda }}} \right)^2} \\
 &= -\frac{E}{6} - \frac{{m_ + ^2}}{4}{\tanh ^2}\frac{z }{2} - \frac{{m_ - ^2}}{4}{\coth ^2}\frac{z }{2}.
\end{split}
\end{equation}
and
\begin{equation}
\begin{split}
\wp ' \left( \tilde{a} \right) &= i {\sqrt{\Delta} \left( {\frac{{m_ + ^2}}{2}{{\left( {\frac{{1 - \lambda }}{{1 + \lambda }}} \right)}^2} - \frac{{m_ - ^2}}{2}{{\left( {\frac{{1 + \lambda }}{{1 - \lambda }}} \right)}^2}} \right)} \\
&= i {\sqrt{\Delta} }\left( \frac{{m_ + ^2}}{2}{\tanh ^2}\frac{z }{2} - \frac{{m_ - ^2}}{2}{\coth ^2}\frac{z }{2} \right) .
\end{split}
\end{equation}
The quantity $\tilde{a}$ as function of $\lambda$ has the property $\tilde{a}\left( 1 / \lambda \right) = \tilde{a} \left( \lambda \right)$.

Substituting the above to the spatial derivative equation of the auxiliary system, we get
\begin{multline}
\frac{d c_i^0 \left( {{\xi^1}} \right)}{d \xi^1}{v_0} + \left[ {\frac{d c_i^ + \left( {{\xi^1}} \right)}{d \xi^1} - ig\left( {{\xi^1}} \right){c_i^ + }\left( {{\xi^1}} \right)} \right]{v_ + }{e^{i\sqrt{\Delta} {\xi^0}}} \\
+ \left[ {\frac{d c_i^ - \left( {{\xi^1}} \right)}{d \xi^1} + ig\left( {{\xi^1}} \right){c_i^ - }\left( {{\xi^1}} \right)} \right]{v_ - }{e^{ - i\sqrt{\Delta} {\xi^0}}} = 0 ,
\end{multline}
implying that
\begin{align}
{c_i^0}\left( {{\xi^1}} \right) &= {c_i^0}\\
{c_i^ \pm }\left( {{\xi^1}} \right) &= {c_i^ \pm }{e^{ \pm i\int {d{\xi^1}g\left( {{\xi^1}} \right)} }}: = {c_i^ \pm }{e^{ \mp i\Phi \left( {{\xi^1}; {\tilde{a}}} \right)}} ,
\end{align}
where the function $\Phi$ is the same quasi-periodic function that appears in the construction of the elliptic strings and it is defined in equation \eqref{eq:elliptic_solutions_lame_phase_def}. Then, 
\begin{equation}
{{\hat \Psi }_i}\left( \lambda  \right) = {c_i^0}{v_0} + {c_i^ + }{v_ + }{e^{i\left( {\sqrt{\Delta} {\xi^0} - \Phi \left( {{\xi^1}; {\tilde{a}}} \right)} \right)}} + {c_i^ - }{v_ - }{e^{ - i\left( {\sqrt{\Delta} {\xi^0} - \Phi \left( {{\xi^1}; {\tilde{a}}} \right)} \right)}} 
\end{equation}
or equivalently
\begin{multline}
{{\hat \Psi }_i}\left( \lambda  \right) = {C_i^1}\left( \lambda  \right)\left[ {\cos \left( {\sqrt{\Delta} {\xi^0} - \Phi \left( {{\xi^1}; {\tilde{a}}} \right)} \right){e_1} + \sin \left( {\sqrt{\Delta} {\xi^0} - \Phi \left( {{\xi^1}; {\tilde{a}}} \right)} \right){e_2}} \right]\\
 + {C_i^2}\left( \lambda  \right)\left[ {- \cos \left( {\sqrt{\Delta} {\xi^0} - \Phi \left( {{\xi^1}; {\tilde{a}}} \right)} \right){e_2} + \sin \left( {\sqrt{\Delta} {\xi^0} - \Phi \left( {{\xi^1}; {\tilde{a}}} \right)} \right){e_1}} \right] + {C_i^3}\left( \lambda  \right){e_3} \\
 : = {C_i^j}\left( \lambda  \right){E_j} ,
\end{multline}
where ${C_i^1} = {c_i^+}+{c_i^-}$, ${C_i^2} = i\left({c_i^+} - {c_i^ -} \right)$ and ${C_i^3} = {c_i^0}$. The vectors ${E_j}$ are defined as
\begin{align}
{E_1} &:= \cos \left( {\sqrt{\Delta} {\xi^0} - \Phi \left( {{\xi^1}; {\tilde{a}}} \right)} \right){e_1} + \sin \left( {\sqrt{\Delta} {\xi^0} - \Phi \left( {{\xi^1}; {\tilde{a}}} \right)} \right){e_2} , \\
{E_2} &:= - \cos \left( {\sqrt{\Delta} {\xi^0} - \Phi \left( {{\xi^1}; {\tilde{a}}} \right)} \right){e_2} + \sin \left( {\sqrt{\Delta} {\xi^0} - \Phi \left( {{\xi^1}; {\tilde{a}}} \right)} \right){e_1} , \\
{E_3} &:= {e_3} 
\end{align}
and they obey $E_i^T {E_j} = \delta_{ij}$ and $E_i \times {E_j} = - \varepsilon_{ijk}E_k$. Notice that as $\lambda \to 0$,
\begin{equation}
\Delta\left( 0 \right) = {x_1} - \wp \left( a \right) = {\ell ^2},\quad \tilde{a}\left( 0 \right) = a
\end{equation}
and thus,
\begin{equation}
{\left. {\sqrt{\Delta} {\xi^0} - \Phi \left( {{\xi^1};\tilde{a}} \right)} \right|_{\lambda  = 0}} = \ell {\xi^0} - \Phi \left( {{\xi^1};a} \right) = \varphi \left( {{\xi^0},{\xi^1}} \right) .
\end{equation}
Therefore,
\begin{equation}
{E_1}\left( 0 \right) = \left( {\begin{array}{*{20}{c}}
{ - {F_2}\cos \varphi }\\
{\sin \varphi }\\
{ - {F_1}\cos \varphi }
\end{array}} \right),\quad {E_2}\left( 0 \right) = \left( {\begin{array}{*{20}{c}}
{ - {F_2}\sin \varphi }\\
{ - \cos \varphi }\\
{ - {F_1}\sin \varphi }
\end{array}} \right),\quad {E_3}\left( 0 \right) = \left( {\begin{array}{*{20}{c}}
{{F_1}}\\
0\\
{ - {F_2}}
\end{array}} \right).
\end{equation}
Additionally, the properties \eqref{eq:dress_ell_e_invertion} imply
\begin{equation}
{E_{1/2}}\left( {1/\lambda } \right) = \theta {E_{1/2}}\left( {\lambda } \right),\quad {E_3}\left( {1/\lambda } \right) =  - \theta {E_3}\left( {\lambda } \right) .
\label{eq:dress_E_vector_inversion_property}
\end{equation}
Finally, notice that the basis vectors $E_i$ have the property
\begin{equation}
{\partial _{0/1}}{E_i} = {\kappa _{0/1}} \times {E_i} .
\end{equation}

Defining the matrices $E$ and $C$ as the matrices comprised by the three columns being the vectors $E_j$ and $C_j$ respectively, the solution can be written in the form
\begin{equation}
{\hat \Psi} \left( \lambda \right) = E C .
\end{equation}

Following the discussion of section \ref{subsec:dressing_involutions}, the solution of the auxiliary system should obey the following constraints
\begin{align}
{\Psi ^\dag }\left( {\bar \lambda } \right)\Psi \left( \lambda  \right) &= I , \\
\overline {\Psi \left( {\bar \lambda } \right)}  &= \Psi \left( \lambda  \right) , \\
\Psi \left( \lambda  \right) &= \Psi \left( 0 \right)\theta \Psi \left( {1/\lambda } \right)\theta .
\end{align}
In terms of the matrix ${\hat \Psi }$, they are written as
\begin{align}
{{\hat \Psi }^T}\left( \lambda  \right)\hat \Psi \left( \lambda  \right) &= I , \label{eq:dress_condition_Psi_hat_orthogonality}\\
\overline {\hat \Psi \left( {\bar \lambda } \right)}  &= \hat \Psi \left( \lambda  \right) ,\label{eq:dress_condition_Psi_hat_reality} \\
\hat \Psi \left( \lambda  \right) &= \theta \hat \Psi \left( {1/\lambda } \right)\theta .\label{eq:dress_condition_Psi_hat_coset}
\end{align}

The reality condition \eqref{eq:dress_condition_Psi_hat_reality} implies that the matrix $C$ obeys the constraint
\begin{equation}
\overline {C\left( {\bar \lambda } \right)}  = C\left( \lambda  \right) .
\label{eq:dress_condition_C_reality}
\end{equation}

The orthogonality condition \eqref{eq:dress_condition_Psi_hat_orthogonality} implies that the matrix $C$ is also orthogonal
\begin{equation}
C ^T \left( \lambda  \right) C \left( \lambda  \right) = I.
\label{eq:dress_condition_C_orthogonality}
\end{equation}

Finally, the coset condition \eqref{eq:dress_condition_Psi_hat_coset} implies that 
\begin{equation}
C \left( 1 / \lambda \right) = \theta C \left( \lambda \right) \theta,
\label{eq:dress_condition_C_coset}
\end{equation}
since the matrix $E$ obeys $E \left( 1 / \lambda \right) =\theta E \left( \lambda \right) \theta $, which is a direct consequence of equation \eqref{eq:dress_E_vector_inversion_property}.

Finally, the solution should obey
\begin{equation}
\hat \Psi \left( 0 \right) = {U^T} = \left( {\begin{array}{*{20}{c}}
{{F_2}\cos \varphi }&{{F_2}\sin \varphi }&{ - {F_1}}\\
{ - \sin \varphi }&{\cos \varphi }&0\\
{{F_1}\cos \varphi }&{{F_1}\sin \varphi }&{{F_2}}
\end{array}} \right) 
\end{equation}
and the matrix $E$ obeys
\begin{equation}
E\left( 0 \right) = \left( {\begin{array}{*{20}{c}}
{ - {F_2}\cos \varphi }&{ - {F_2}\sin \varphi }&{{F_1}}\\
{\sin \varphi }&{ - \cos \varphi }&0\\
{ - {F_1}\cos \varphi }&{ - {F_1}\sin \varphi }&{ - {F_2}}
\end{array}} \right) .
\end{equation}
It follows that the coefficients matrix should obey
\begin{equation}
C\left( 0 \right) = \left( {\begin{array}{*{20}{c}}
{ - 1}&0&0\\
0&{ - 1}&0\\
0&0&{ - 1}
\end{array}} \right) = - I.
\label{eq:dress_condition_C_initial}
\end{equation}

Thus, it is simple to satisfy all the conditions \eqref{eq:dress_condition_C_reality}, \eqref{eq:dress_condition_C_orthogonality}, \eqref{eq:dress_condition_C_coset} and \eqref{eq:dress_condition_C_initial}, selecting
\begin{equation}
C\left( \lambda  \right) = - I ,
\end{equation}
implying that the solution of the auxiliary system that obeys all the appropriate involutions and the initial condition is
\begin{equation}
\Psi_{ij} \left( \lambda  \right) =  - E_j^i.
\label{eq:dressed_string_auxiliary_solution}
\end{equation}

\subsection{The Dressed Solution in the Case of Two Poles}
\label{subsec:dressed_strings_two_poles}
As analysed in section \ref{sec:dress_review}, the simplest possible dressing factor has two poles lying on the unit circle at positions complex conjugate to each other. In this case, the dressed solution is
\begin{equation}
f' = \chi \left( 0 \right)\Psi \left( 0 \right) ,
\end{equation}
where $\chi \left( \lambda \right)$ is given by equations \eqref{eq:dress_review_dressing_factor_two_conjugate_poles} and \eqref{eq:dress_review_dressing_factor_two_conjugate_poles_projector}.
The constant vector $p$ obeys ${p^T}p = 0$, $\bar p = \theta p$ and thus, it may be parametrized in terms of two real numbers $a$ and $b$ as
\begin{equation}
p = \left( {\begin{array}{*{20}{c}}
{a\cos b}\\
{a\sin b}\\
{ia}
\end{array}} \right) .
\label{eq:dressed_strings_projector}
\end{equation}
We also define
\begin{equation}
\lambda_1 = e^{i \theta_1}.
\end{equation}

In order to visualize and understand the behaviour of the dressed solution, we would like to find the unit vector $X'$ that corresponds to the coset element $f'$ through the mapping \eqref{eq:dressing_mapping_coset}. For this purpose we define
\begin{equation}
f' = \theta U\theta \hat f'{U^T} .
\end{equation}
Then
\begin{equation}
\hat f' = \theta \left( I - 2 {{\hat X}'} {{\hat X}^{\prime T}} \right) ,
\end{equation}
where
\begin{equation}
X ' = U {\hat X} ' .
\end{equation}
in a similar manner to the definitions we used to solve the auxiliary system. Then,
\begin{equation}
\hat f' = I - \frac{{{\lambda _1} - 1/{\lambda _1}}}{{{\lambda _1}}}\frac{{\theta \hat \Psi \left( {{\lambda _1}} \right)\theta e{e^T}\theta {{\hat \Psi }^T}\left( {{\lambda _1}} \right)}}{{{e^T}\theta {{\hat \Psi }^T}\left( {{\lambda _1}} \right) \theta \hat \Psi \left( {{\lambda _1}} \right)\theta e}} - \frac{{1/{\lambda _1} - {\lambda _1}}}{{1/{\lambda _1}}}\frac{{\hat \Psi \left( {{\lambda _1}} \right)\theta e{e^T}\theta {{\hat \Psi }^T}\left( {{\lambda _1}} \right)\theta }}{{{e^T}\theta {{\hat \Psi }^T}\left( {{\lambda _1}} \right) \theta \hat \Psi \left( {{\lambda _1}} \right)\theta e}}
\end{equation}
or
\begin{equation}
\hat f' = I - \frac{{{\lambda _1} - 1/{\lambda _1}}}{{{\lambda _1}}}\frac{{{X_-}X_+^T}}{{X_+^T{X_-}}} - \frac{{1/{\lambda _1} - {\lambda _1}}}{{1/{\lambda _1}}}\frac{{{X_+}X_-^T}}{{X_+^T{X_-}}} ,
\end{equation}
where
\begin{equation}
{X_+} = \hat \Psi \left( {{\lambda _1}} \right)\theta e ,\quad {X_-} = \theta \hat \Psi \left( {{\lambda _1}} \right)\theta e .
\label{eq:dressed_string_Xpm_def}
\end{equation}
The vectors $X_\pm$ obey the property $X_\pm^T (X_\pm)  = 0$, they are complex conjugate to each other and they are eigenvectors of $\hat f '$ since
\begin{equation}
\hat f'{X_\pm} = e^{\pm 2 i \theta_1}{X_\pm} .
\end{equation}
In section \ref{subsec:mapping_X_f}, we showed that the orthogonal matrix $f = \left( I - 2{X_0}X_0^T \right) \left( {I - 2X{X^T}} \right)$ has three eigenvectors, the vector $\chi_0 = X_0 \times X$ with eigenvalue equal to one, and the vectors $\chi_\pm = { - {e^{\pm i\theta_1 }}{X_0} + X} $ with eigenvalues $e^{\pm 2 i a}$, where $a$ is the angle between $X_0$ and $X$. It follows that the vectors $X_\pm$ are actually proportional to the eigenvectors $\hat \chi_\pm ' = { - {e^{\pm i\theta_1 }}{X_0} + \hat X '} $ and furthermore that the vector $\hat X '$ forms an angle $\theta_1$ with $X_0$. The proportionality constant can be fixed so that their inner product matches that of $\hat \chi_\pm ' $, which equals $2 \sin^2 \theta_1$. Thus,
\begin{equation}
\hat \chi_\pm ' = - {e^{ \pm i\theta_1 }}{X_0} + \hat X ' = \sqrt {\frac{2}{{X_ + ^T{X_ - }}}} \sin \theta_1 {X_ \pm } 
\end{equation}
and finally,
\begin{equation}
\hat X ' = \sqrt {\frac{1}{{2X_ + ^T{X_ - }}}} \sin \theta_1 \left( {{X_ + } + {X_ - }} \right) + \cos \theta_1 {X_0} := \sin \theta_1 {X_1} + \cos \theta_1 {X_0}.
\label{eq:dressed_solution_x_hat_prime}
\end{equation}
Thus, the dressed string solution is
\begin{equation}
X ' = U \hat X ' ,
\label{eq:dressed_solution_final}
\end{equation}
where $\hat X '$ is given by \eqref{eq:dressed_solution_x_hat_prime}.

It is easy to show that the vector $X_1$ is a unit vector, which is perpendicular to $X_0$, due to the fact that $X_- = \theta X_+$. Thus, the equation \eqref{eq:dressed_solution_x_hat_prime} implies that the arc connecting the endpoints of the vectors $X_0$ and $\hat X '$ is equal to $\theta_1$. Since the seed solution is given by $X = U \hat X = U X_0$ and the dressed solution is given by $X' = U \hat X'$, this property is transferred to the points of the seed and dressed solutions that correspond to the same worldsheet parameters $\xi^{0/1}$. In other words, \emph{the dressed string solution can be visualized as being drawn by a point in the circumference of an epicycle of arc radius $\theta_1$, which moves so that its center lies on the seed string solution}. 

This statement provides a nice geometric visualization of the action of the dressing on the shape of the string. It is a general property that follows from the equation \eqref{eq:dressed_solution_x_hat_prime}, which is the outcome of the form of the dressing factor in the case it has only two poles \eqref{eq:dress_review_dressing_factor_two_conjugate_poles} as well as the mapping \eqref{eq:dressing_mapping_coset} between unit vectors and elements of the coset $\mathrm{SO}(3)/\mathrm{SO}(2)$. It follows that the epicycle picture is \emph{not} a specific property of the dressed elliptic solutions, but a \emph{generic} property that holds whenever the simplest dressing factor is adopted. This interesting property of the dressing method deserves further investigation in the case of strings propagating on other symmetric spaces or in the case of a more complicated dressing factor. A further implication of the above is the fact that at the limit $\theta_1 \to 0$ the dressed solution tends to the seed, whereas as $\theta_1 \to \pi$ the dressed solution tends to the reflection of the seed with respect to the origin of the enhanced space.

\begin{figure}[ht]
\begin{center}
\begin{picture}(80,78.5)
\put(0,4){\includegraphics[width = 0.35\textwidth]{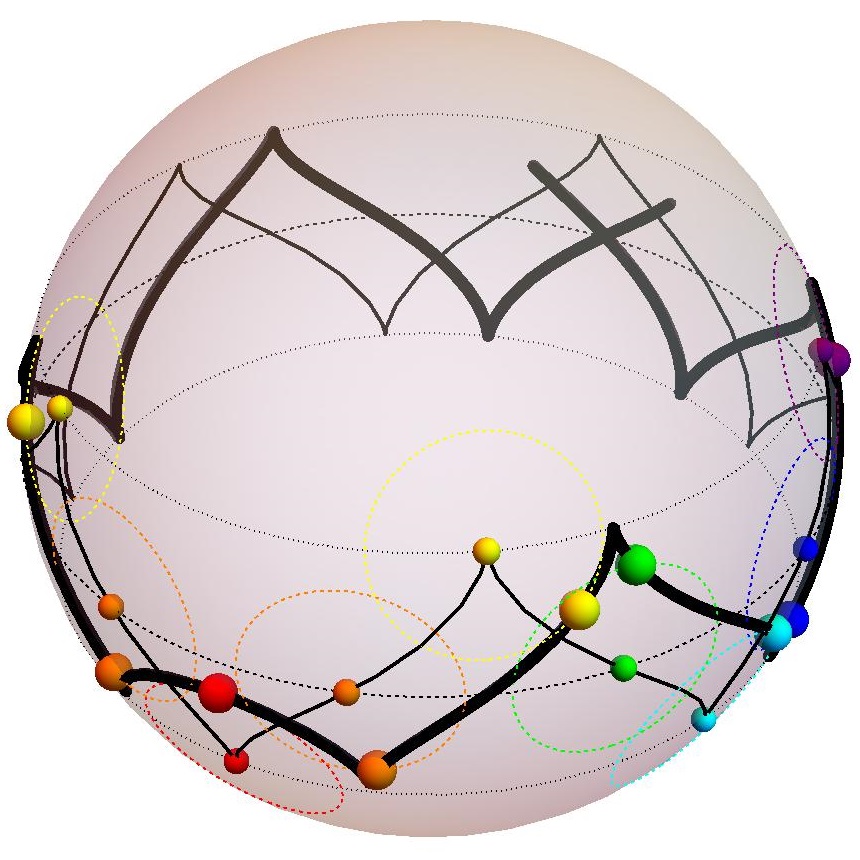}}
\put(45,4){\includegraphics[width = 0.35\textwidth]{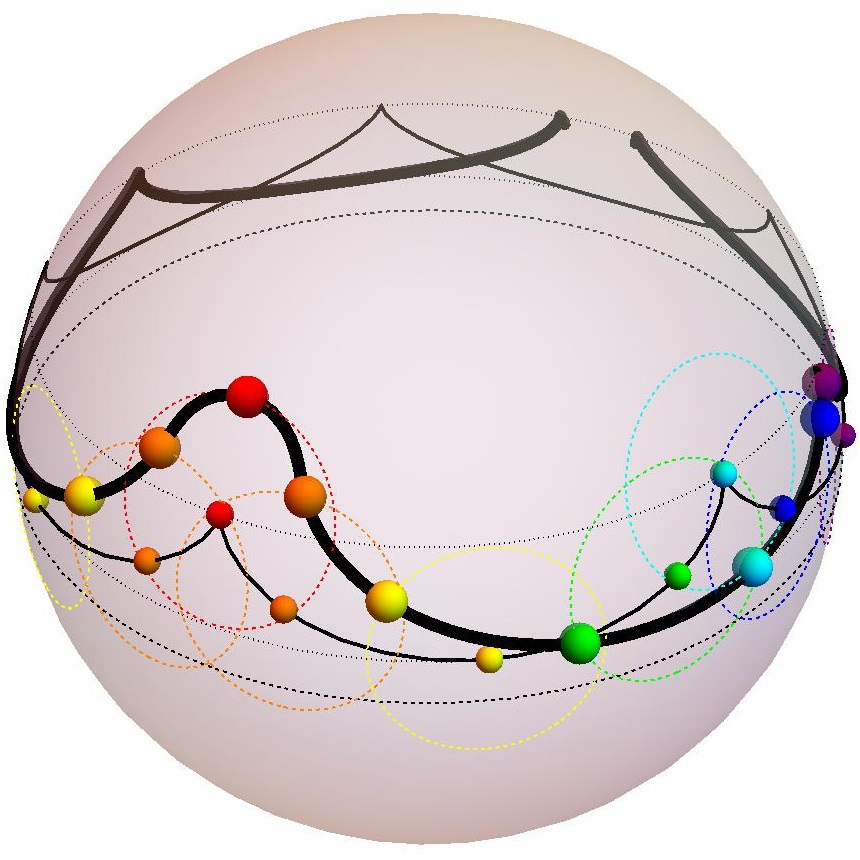}}
\put(0,43.5){\includegraphics[width = 0.35\textwidth]{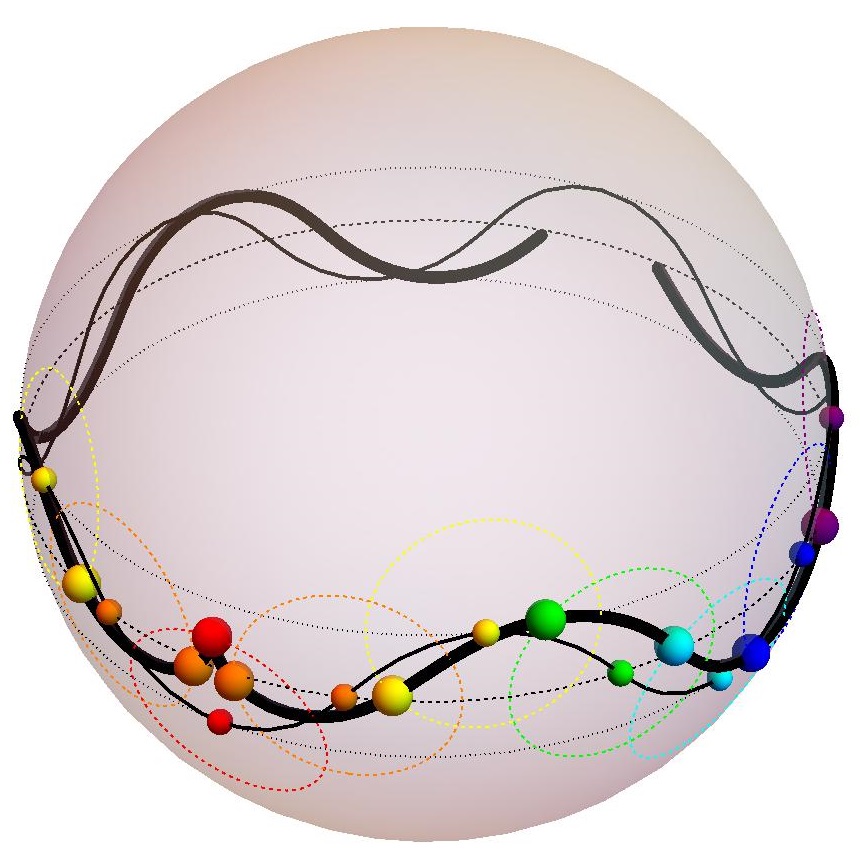}}
\put(45,43.5){\includegraphics[width = 0.35\textwidth]{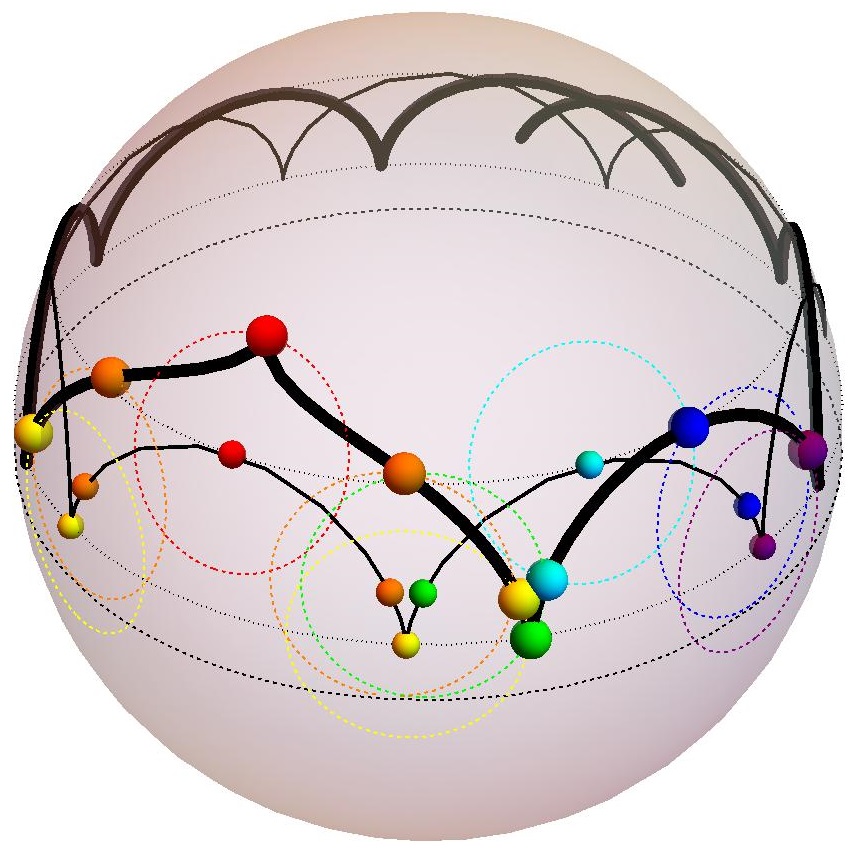}}
\put(7.5,42.25){seed with static}
\put(4,39.5){oscillating counterpart}
\put(52.5,42.25){seed with static}
\put(50,39.5){rotating counterpart}
\put(-2.5,2.75){seed with translationally invariant}
\put(4,0){oscillating counterpart}
\put(42.5,2.75){seed with translationally invariant}
\put(50,0){rotating counterpart}
\end{picture}
\end{center}
\vspace{-10pt}
\caption{The dressed elliptic string solutions}
\vspace{5pt}
\label{fig:dressed_strings_static_circles}
\end{figure}
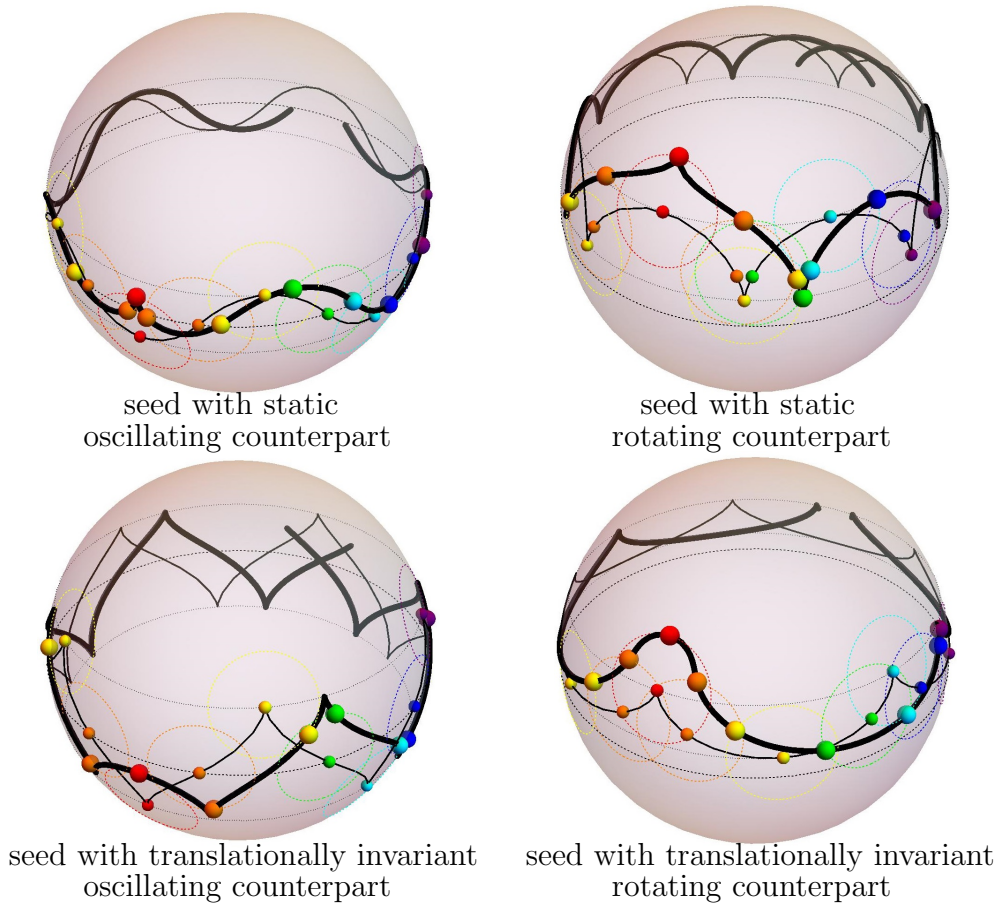

In figure \ref{fig:dressed_strings_static_circles}, four representative dressed elliptic string solutions are depicted. In these plots, the dressed string solutions are depicted with a thick black line, whereas the seed solutions are depicted with a thin one. In the top row, the seed solution has a translationally invariant elliptic Pohlmeyer counterpart, whereas in the bottom row it has a static one. On the left column the seed solution has an oscillating counterpart with $E=\mu^2 / 10$ and $a$ selected so that $n=10$, whereas on the right column the seed solution has a rotating counterpart with $E=6\mu^2 /5$ and $a$ selected so that $n=7$. In all cases the pair of poles of the dressing factor lies at $\lambda = e^{\pm i\frac{\pi}{12}}$. Large spheres are points of the dressed solution, whereas small spheres are points of the seed solution. Spheres with the same color correspond to the same worldsheet coordinates $\xi_0$ and $\xi_1$ and they are connected via an epicycle plotted with the same color, too.

Our analysis focused on seed solutions being elliptic string solutions with static Pohlmeyer counterparts. It is trivial to show that had we used elliptic strings with translationally invariant Pohlmeyer counterparts as seed solutions, we would have resulted in dressed string solutions that can be acquired from the ones presented here after the trivial operation ${\xi ^0} \leftrightarrow {\xi ^1}$.

\setcounter{equation}{0}
\section{The Sine-Gordon Equation Counterparts}
\label{sec:Elliptic_kinks}

The elliptic string solutions presented in section \ref{sec:review} can be naturally classified with respect to their Pohlmeyer counterparts. Furthermore, in \cite{part1} it was also shown that many of the properties of these solutions are connected to the properties of their corresponding sine-Gordon counterparts. For example, the number of spikes equals the topological number in the sine-Gordon theory. For these reasons, we proceed to specify in this section the sine-Gordon equation counterparts of the dressed elliptic string solutions, which are obtained in section \ref{sec:dressed_strings}.

\subsection{\Backlund Transformations}
\label{subsec:Backlund}

The sine-Gordon equation \eqref{eq:Pohlmeyer_SGequation} possesses the \Backlund transformations
\begin{align}
{\partial _ + }\frac{{\varphi  + \tilde \varphi }}{2} &= a\mu \sin \frac{{\varphi  - \tilde \varphi }}{2} , \label{eq:SG_dress_Backlund_1}\\
{\partial _ - }\frac{{\varphi  - \tilde \varphi }}{2} &= \frac{1}{a}\mu \sin \frac{{\varphi  + \tilde \varphi }}{2} , \label{eq:SG_dress_Backlund_2}
\end{align}
connecting pairs of solutions. As described in the introduction, they can be used for the construction of new solutions from a seed one. Their merit is the fact that this is achieved via solving a pair of first order differential equations, instead of the original second order one. The usual application of these transformations is the construction of the kink solutions, using the vacuum $\varphi = 0$ as seed.

A nice property of the \Backlund transformations is the fact their iterative use does not require further solving of differential equations. Multi-kink solutions can be acquired from the single-kink ones algebraically using the Bianchi permutability theorem. If $\varphi_1$ is connected to the seed $\varphi$ through a \Backlund transformation with parameter $a_1$ and $\varphi_2$ is connected to the same seed $\varphi$ through a \Backlund transformation with parameter $a_2$, then a new solution $\varphi_{12}$ that is connected to $\varphi_1$ through a \Backlund transformation with parameter $a_2$ (or equivalently to $\varphi_2$ through a \Backlund transformation with parameter $a_1$) will be given by
\begin{equation}
\tan \frac{{{\varphi _{12}} - \varphi }}{4} = \frac{{{a_1} + {a_2}}}{{{a_1} - {a_2}}}\tan \frac{{{\varphi _1} - {\varphi _2}}}{4} .
\label{eq:SG_addition_theorem}
\end{equation}

\subsection{Virasoro Constraints}

A basic ingredient of the Pohlmeyer reduction is the fact that the energy momentum tensor can be set constant, with obvious consequences for the form of the Virasoro constraints. In the following, as a first step towards the specification of the Pohlmeyer counterparts of the dressed solutions discovered in section \ref{sec:dressed_strings}, we show explicitly that  they obey the Virasoro constraints as expected by the analysis in section \ref{subsec:dress_review_Virasoro}.

We have shown that the dressed solution can be written as
\begin{equation}
X' = U\hat X' =U \left( {X_1}\sin \theta_1  + {X_0}\cos \theta_1 \right) .
\label{eq:app_VP_hatprime}
\end{equation}
The vectors $X_0$ and $X_1$ are unit vectors, orthogonal to each other, thus the vectors $\left\{X_0 , X_1 , X_0 \times X_1 \right\}$ form an orthonormal basis.

By the definition \eqref{eq:dressed_string_Xpm_def} of the vector $X_+$, we have
\begin{equation}
{X_ + } = \hat \Psi \left( {{\lambda _1}} \right)\theta p =  - {\left( {\theta p} \right)^i}{E_i}
\end{equation}
and we have already shown that ${\partial _{0/1}}{E_i} = {{\kappa }_{0/1}} \times {E_i} $. Therefore,
\begin{equation}
{\partial _{0/1}}{X_ + } = {{\kappa }_{0/1}} \times {X_ + } .
\label{eq:app_VP_Xp_der}
\end{equation}
In a similar manner, ${X_ - } = \theta {X_ + }$, and, thus,
\begin{equation}
{\partial _{0/1}}{X_ - } = \theta \left( {{{\kappa }_{0/1}} \times {X_ + }} \right) = \left( { - \theta {{\kappa }_{0/1}}} \right) \times {X_ - } .
\label{eq:app_VP_Xm_der}
\end{equation}

The third element of the vector $X_1$ vanishes, as it is perpendicular to $X_0$. Thus, the third element of its derivative also vanishes. Since $X_1$ is a constant norm vector, its derivatives are perpendicular to itself. The above imply that the derivatives of $X_1$ are perpendicular to both $X_0$ and $X_1$, thus parallel to $X_0 \times X_1$,
\begin{equation}
{\partial _{0/1}}{X_1} = {c_{0/1}}{X_0} \times {X_1} .
\label{eq:app_der_X1}
\end{equation}
The formulae \eqref{eq:app_VP_Xp_der} and \eqref{eq:app_VP_Xm_der} that provide the derivatives of $X_\pm$, can be used to calculate the coefficients $c_{0/1}$. It is a matter of algebra to show that
\begin{multline}
{\partial _i}{X_1} \cdot \left( {{X_0} \times {X_1}} \right) = \left( {\partial _i}{X_1} \right) ^T \left( {{X_0} \times {X_1}} \right) = \frac{1}{{2{X_ + ^T} {X_ - }}}\left[ {\left( {{X_0^T} \left( {{{\kappa }_i} + \left( { - \theta {{\kappa }_i}} \right)} \right)} \right)} \right.\left( {{X_ + ^T} {X_ - }} \right)\\
\left. { - \left( {{X_0^T} {X_ + }} \right)\left( {{{\kappa }_i^T} \left( {{X_ + } + {X_ - }} \right)} \right) - \left( {{X_0^T} {X_ - }} \right)\left( {\left( { - \theta {{\kappa }_i}} \right)^T  \left( {{X_ + } + {X_ - }} \right)} \right)} \right] .
\label{eq:app_VP_equation_c}
\end{multline}
Recalling the definitions \eqref{eq:defs_kappa_1} and \eqref{eq:defs_kappa_2} of the $\kappa_i$ in terms of the real vectors $k_i$, it is obvious that
\begin{align}
{{\kappa }_i^T} {X_0} = \left( { - \theta {{\kappa }_i}} \right)^T {X_0} &=  - {X_0} \cdot {{k}_i} , \label{eq:app_VP_kappadotX0}\\
{{\kappa }_i^T} \left( {{X_ + } + {X_ - }} \right) = - \left( { - \theta {{\kappa }_i}} \right)^T \left( {{X_ + } + {X_ - }} \right) &= i\left( {\cot \theta_1 {{k}_i} - \csc \theta_1 {{k}_{\bar i}}} \right) \cdot \left( {{X_ + } + {X_ - }} \right) , \label{eq:app_VP_kappadotXpm}
\end{align}
where $\bar i = 0$ when $i = 1$ and vice versa.
Equations \eqref{eq:app_VP_kappadotX0} and \eqref{eq:app_VP_kappadotXpm}, together with the property $\sqrt {2{X_ +^T } {X_ - }}  =  - i{X_0^T} \left( {{X_ + } - {X_ - }} \right)$, which is a direct consequence of the properties of the vector $p$, allow us to write equation \eqref{eq:app_VP_equation_c} as, 
\begin{equation}
{\partial _i}{X_1} \cdot \left( {{X_0} \times {X_1}} \right) =  - {X_0} \cdot {{k}_i} + \left( {\cot \theta_1 {{k}_i} - \csc \theta_1 {{k}_{\bar i}}} \right) \cdot {X_1} ,
\end{equation}
implying
\begin{equation}
{c_i} =  - {X_0} \cdot {{k}_i} + \left( {\cot \theta_1 {{k}_i} - \csc \theta_1 {{k}_{\bar i}}} \right) \cdot {X_1} .
\label{eq:app_der_x1_cs}
\end{equation}

Equation \eqref{eq:app_VP_hatprime} implies that
\begin{equation}
{\partial _i}X' = \left( {{\partial _i}U} \right)\hat X' + \sin \theta_1 U\left( {{\partial _i}{X_1}} \right) .
\end{equation}
A direct consequence of the above is
\begin{multline}
\left( {{\partial _i}X'} \right) \cdot \left( {{\partial _j}X'} \right) = \hat X{'^T}\left( {{\partial _i}{U^T}} \right)U{U^T}\left( {{\partial _j}U} \right)\hat X' + {\sin ^2}\theta_1 \left( {{\partial _i}X_1^T} \right)\left( {{\partial _j}{X_1}} \right)\\
 + \sin \theta_1 \left[ {\hat X{'^T}\left( {{\partial _i}{U^T}} \right)U\left( {{\partial _j}{X_1}} \right) + \left( {{\partial _i}X_1^T} \right){U^T}\left( {{\partial _j}U} \right)\hat X'} \right]\\
 = \left( {{U^T}\left( {{\partial _i}U} \right)\hat X'} \right) \cdot \left( {{U^T}\left( {{\partial _j}U} \right)\hat X'} \right) + {\sin ^2}\theta_1 \left( {{\partial _i}{X_1}} \right) \cdot \left( {{\partial _j}{X_1}} \right)\\
 + \sin \theta_1 \left[ {\left( {{U^T}\left( {{\partial _i}U} \right)\hat X'} \right) \cdot \left( {{\partial _j}{X_1}} \right) + \left( {{\partial _i}X_1^T} \right) \cdot \left( {{U^T}\left( {{\partial _j}U} \right)\hat X'} \right)} \right] .
\end{multline}
We remind the reader that we have defined the vectors $k_i$ so that ${U^T}\left( {{\partial _j}U} \right) = k_j^i{T_i}$. This implies that ${U^T}\left( {{\partial _i}U} \right)\hat X' = {{k}_i} \times \hat X'$. Taking advantage of this and the form of the derivatives of the vector $X_1$ \eqref{eq:app_der_X1}, we find
\begin{multline}
\left( {{\partial _i}X'} \right) \cdot \left( {{\partial _j}X'} \right) = \left( {\sin \theta_1 \left( {{{k}_i} + {c_i}{X_0}} \right) \times {X_1} + \cos \theta_1 {{k}_i} \times {X_0}} \right) \\
\cdot \left( {\sin \theta_1 \left( {{{k}_j} + {c_j}{X_0}} \right) \times {X_1} + \cos \theta_1 {{k}_j} \times {X_0}} \right) .
\end{multline}

Putting everything together, it is now a matter of simple algebra to show that
\begin{align}
\left( {{\partial _0}X'} \right) \cdot \left( {{\partial _1}X'} \right) &= \left( {{{k}_0} \times {X_0}} \right) \cdot \left( {{{k}_1} \times {X_0}} \right) \nonumber\\
&= \left( {{\partial _0}X} \right) \cdot \left( {{\partial _1}X} \right) , \label{eq:app_Vir1}\\
\left( {{\partial _0}X'} \right) \cdot \left( {{\partial _0}X'} \right) + \left( {{\partial _1}X'} \right) \cdot \left( {{\partial _1}X'} \right) &= {\left| {{{k}_0} \times {X_0}} \right|^2} + {\left| {{{k}_1} \times {X_0}} \right|^2} \nonumber\\
&= \left( {{\partial _0}X} \right) \cdot \left( {{\partial _0}X} \right) + \left( {{\partial _1}X} \right) \cdot \left( {{\partial _1}X} \right) , \label{eq:app_Vir2}
\end{align}
implying that the dressed solution satisfies the Virasoro constraints as long as the undressed solution does so.

\subsection{Dressing vs \Backlund Transformation}

Let us now study the connection of the Pohlmeyer field corresponding to the dressed solution to that of the seed. In exactly the same way that we derived \eqref{eq:app_Vir1} and \eqref{eq:app_Vir2}, we find
\begin{multline}
\left( {{\partial _0}X'} \right) \cdot \left( {{\partial _0}X'} \right) - \left( {{\partial _1}X'} \right) \cdot \left( {{\partial _1}X'} \right) \\
= {\left| {{{k}_0} \times {X_0}} \right|^2} - {\left| {{{k}_1} \times {X_0}} \right|^2} - 2\left[ {{{\left( {{{k}_0} \cdot {X_1}} \right)}^2} - {{\left( {{{k}_1} \cdot {X_1}} \right)}^2}} \right]\\
= \left( {{\partial _0}X} \right) \cdot \left( {{\partial _0}X} \right) - \left( {{\partial _1}X} \right) \cdot \left( {{\partial _1}X} \right) - 2\left[ {{{\left( {{{k}_0} \cdot {X_1}} \right)}^2} - {{\left( {{{k}_1} \cdot {X_1}} \right)}^2}} \right] .
\label{eq:app_Pohlmayer_field}
\end{multline}

Taking advantage of the fact that $k_0^2$ and $k_0^3k_1^1 - k_1^3k_0^1 = 0$, we may write the derivatives of the $1$ and $2$ components of the vectors $k_0$ and $k_1$ \eqref{eq:dressed_string_k1_der} and \eqref{eq:dressed_string_k2_der} as
\begin{align}
{\partial _1}k_0^1 =  - k_0^3k_1^2 + k_1^3k_0^2,&\quad {\partial _1}k_0^2 = k_0^3k_1^1 - k_1^3k_0^1 , \\
{\partial _1}k_1^1 =  - k_0^3k_0^2 + k_1^3k_1^2,&\quad {\partial _1}k_1^2 = k_0^3k_0^1 - k_1^3k_1^1 .
\end{align}
We remind the reader that ${\partial _1} k_{0/1} = 0$.
The above imply that the perpendicular to $X_0$ part of the the derivatives of $k_0$ and $k_1$ can be written as
\begin{align}
{\left( {{\partial _1}{{k}_0}} \right)_ \bot } &= \left( {{{k}_0} \cdot {X_0}} \right){X_0} \times {{k}_1} - \left( {{{k}_1} \cdot {X_0}} \right){X_0} \times {{k}_0} , \\
{\left( {{\partial _1}{{k}_1}} \right)_ \bot } &= \left( {{{k}_0} \cdot {X_0}} \right){X_0} \times {{k}_0} - \left( {{{k}_1} \cdot {X_0}} \right){X_0} \times {{k}_1} .
\end{align}
Defining
\begin{equation}
{{k}_ \pm } = {{k}_1} \pm {{k}_0} ,
\end{equation}
the above relations can be written in a shorthand notation as
\begin{equation}
{\left( {{\partial _1}{{k}_ \pm }} \right)_ \bot } =  - \left( {{{k}_ \mp } \cdot {X_0}} \right){X_0} \times {{k}_ \pm } .
\label{eq:app:der_of_kpm}
\end{equation}

We remind the reader that the vectors $X_0$, $X_1$ and $X_0 \times X_1$ form an orthonormal basis. We may project the above relations in the directions of the last two vectors of this basis to yield
\begin{align}
\left( {{\partial _1}{{k}_ \pm }} \right) \cdot {X_1} &= \left( {{{k}_ \mp } \cdot {X_0}} \right)\left( {{{k}_ \pm } \cdot \left( {{X_0} \times {X_1}} \right)} \right) , \\
\left( {{\partial _1}{{k}_ \pm }} \right) \cdot \left( {{X_0} \times {X_1}} \right) &=  - \left( {{{k}_ \mp } \cdot {X_0}} \right)\left( {{{k}_ \pm } \cdot {X_1}} \right) .
\end{align}
In the following we adopt the notation
\begin{equation}
v \cdot {X_0} \equiv {v^a},\quad v \cdot {X_1} \equiv {v^b},\quad v \cdot \left( {{X_0} \times {X_1}} \right) \equiv {v^c} .
\end{equation}
In this notation, appropriately combining the equations \eqref{eq:app_der_X1} and \eqref{eq:app_der_x1_cs} yields
\begin{align}
{\partial _ + }{X_1} = \left( { - k_ + ^a - \tan \frac{\theta_1 }{2}k_ + ^b} \right){X_0} \times {X_1},&\quad {\partial _ + } \left( {X_0} \times {X_1} \right) = \left( {k_ + ^a + \tan \frac{\theta_1 }{2}k_ + ^b} \right){X_1} , \\
{\partial _ - }{X_1} = \left( { - k_ - ^a + \cot \frac{\theta_1 }{2}k_ - ^b} \right){X_0} \times {X_1},&\quad {\partial _ - } \left( {X_0} \times {X_1} \right) = \left( {k_ - ^a - \cot \frac{\theta_1 }{2}k_ - ^b} \right){X_1} .
\end{align}
The above equations and \eqref{eq:app:der_of_kpm} imply that
\begin{align}
{\partial _ - }k_ + ^b &= k_ + ^ck_ - ^b\cot \frac{\theta_1 }{2} ,\label{eq:app_k_Backlund_1}\\
{\partial _ + }k_ - ^b &=  - k_ - ^ck_ + ^b\tan \frac{\theta_1 }{2} . \label{eq:app_k_Backlund_2}
\end{align}

The Virasoro constraints \eqref{eq:app_Vir1} and \eqref{eq:app_Vir2} directly imply that
\begin{equation}
m_ \pm ^2 = \left| {{{k}_ \pm } \times {X_0}} \right|^2 = {\big( {k_ \pm ^b} \big)^2} + {\left( {k_ \pm ^c} \right)^2} .
\label{eq:app_Virasoro_ks}
\end{equation}
In a similar manner, equation \eqref{eq:app_Pohlmayer_field} and the definition of the Pohlmeyer field \eqref{eq:Pohlmeyer_sG_field_definition} imply that
\begin{align}
{m_ + }{m_ - }\cos \varphi  &= \left( {{{k}_ + } \times {X_0}} \right) \cdot \left( {{{k}_ - } \times {X_0}} \right) = k_ + ^bk_ - ^b + k_ + ^ck_ - ^c , \label{eq:app_Pohlmeyer_naked_ks}\\
{m_ + }{m_ - }\cos \tilde \varphi  &= \left( {{{k}_ + } \times {X_0}} \right) \cdot \left( {{{k}_ - } \times {X_0}} \right) - 2\left( {{{k}_ + } \cdot {X_1}} \right)\left( {{{k}_ - } \cdot {X_1}} \right) =  - k_ + ^bk_ - ^b + k_ + ^ck_ - ^c .\label{eq:app_Pohlmeyer_dressed_ks}
\end{align}
It is a direct consequence of \eqref{eq:app_Virasoro_ks}, \eqref{eq:app_Pohlmeyer_naked_ks} and \eqref{eq:app_Pohlmeyer_dressed_ks} that
\begin{align}
{m_ + }{m_ - }\sin \varphi  &=  - k_ + ^bk_ - ^c + k_ + ^ck_ - ^b , \label{eq:app_Pohlmeyer_sin_naked_ks}\\
{m_ + }{m_ - }\sin \tilde \varphi  &= k_ + ^bk_ - ^c + k_ + ^ck_ - ^b , \label{eq:app_Pohlmeyer_sin_dressed_ks}
\end{align}
up to an overall sign which corresponds to the freedom of reflection of the Pohlmeyer field. The equations \eqref{eq:app_Virasoro_ks}, \eqref{eq:app_Pohlmeyer_naked_ks}, \eqref{eq:app_Pohlmeyer_dressed_ks},\eqref{eq:app_Pohlmeyer_sin_naked_ks} and \eqref{eq:app_Pohlmeyer_sin_dressed_ks} imply that
\begin{align}
k_ + ^b =  - {m_ + }\sin \frac{{\varphi  - \tilde \varphi }}{2},\quad k_ + ^c = {m_ + }\cos \frac{{\varphi  - \tilde \varphi }}{2} ,\\
k_ - ^b = {m_ - }\sin \frac{{\varphi  + \tilde \varphi }}{2},\quad k_ - ^c = {m_ - }\cos \frac{{\varphi  + \tilde \varphi }}{2} .
\end{align}

Substituting the above in \eqref{eq:app_k_Backlund_1} and \eqref{eq:app_k_Backlund_2} yields
\begin{align}
{\partial _ - }\frac{{\varphi  - \tilde \varphi }}{2} &=  - {m_ - }\cot \frac{\theta_1 }{2}\sin \frac{{\varphi  + \tilde \varphi }}{2} , \\
{\partial _ + }\frac{{\varphi  + \tilde \varphi }}{2} &= {m_ + }\tan \frac{\theta_1 }{2}\sin \frac{{\varphi  - \tilde \varphi }}{2} ,
\end{align}
which are the usual \Backlund transformations \eqref{eq:SG_dress_Backlund_1} and \eqref{eq:SG_dress_Backlund_2} with parameter
\begin{equation}
a = \sqrt { - \frac{{{m_ + }}}{{{m_ - }}}} \tan \frac{\theta_1 }{2} .
\label{eq:dress_vs_Backlund}
\end{equation}
It follows that the dressed string solutions obtained in section \ref{sec:dressed_strings} have Pohlmeyer counterparts that are connected to the elliptic solutions of the sine-Gordon equation presented in section \ref{sec:review} via a single \Backlund transformation with parameter determined by the position of the poles of the dressing factor.
\pagebreak

\subsection{\Backlund Transformation of Elliptic Solutions}
\label{subsec:Backlund_elliptic}

The last step towards obtaining the Pohlmeyer counterparts of the dressed elliptic string solutions of section \ref{sec:dressed_strings} is the application of a \Backlund transformation to the elliptic solutions of the sine-Gordon equation \eqref{eq:elliptic_solution}. Such solutions have been studied in the past \cite{Jaworski:1982ag,Zagrodzinski:1982gc,Zagrodzinski:1984cn,Jaworski:1987} in a different context and language.

In general, a much wider class of solutions of the sine-Gordon equation can be expressed in terms of hyperelliptic functions \cite{Kotlarov:2014yxa,Kozel}. Such solutions can be classified in terms of the genus of the relevant torus. The elliptic solutions that we have studied in section \ref{sec:review} are the simple case of genus-one solutions. Pairs of solutions connected via a \Backlund transformation are characterized by genuses whose difference equals one. This extra hole in the relevant torus is a degenerate one meaning that one of the corresponding periods is infinite. Therefore, the solutions that we are going to construct applying a \Backlund transformation to elliptic solutions are degenerate cases of genus two solutions of the sine-Gordon equation. In a different approach one may find other genus two solutions via separation of variables \cite{LAMB:1971zz,Osborne:1978ii}.

The technical advantage of using an elliptic solution as seed is the fact that they depend solely on either the space-like or time-like coordinate. Writing down the \Backlund transformations \eqref{eq:SG_dress_Backlund_1} and \eqref{eq:SG_dress_Backlund_2} in terms of the worldsheet coordinates $\xi^0$ and $\xi^1$ yields
\begin{align}
{\partial _1}\frac{\varphi }{2} + {\partial _0}\frac{{\tilde \varphi }}{2} &= \frac{\mu }{2}\left( {a + \frac{1}{a}} \right)\sin \frac{\varphi }{2}\cos \frac{{\tilde \varphi }}{2} - \frac{\mu }{2}\left( {a - \frac{1}{a}} \right)\cos \frac{\varphi }{2}\sin \frac{{\tilde \varphi }}{2} , \label{eq:sg_backlund_0} \\
{\partial _0}\frac{\varphi }{2} + {\partial _1}\frac{{\tilde \varphi }}{2} &= \frac{\mu }{2}\left( {a - \frac{1}{a}} \right)\sin \frac{\varphi }{2}\cos \frac{{\tilde \varphi }}{2} - \frac{\mu }{2}\left( {a + \frac{1}{a}} \right)\cos \frac{\varphi }{2}\sin \frac{{\tilde \varphi }}{2} . \label{eq:sg_backlund_1}
\end{align}

Without loss of generality, we start our analysis considering that $\varphi$ is a translationally invariant elliptic solution of the sine-Gordon equation as given by equation \eqref{eq:elliptic_solution_phi_0}. Equation \eqref{eq:elliptic_solution} directly implies that
\begin{align}
{\cos ^2}\frac{\varphi }{2} &= \frac{1}{{{\mu ^2}}}\left( {{x_2} - \wp \left( {{\xi ^0} + {\omega _2}} \right)} \right) , \\
{\sin ^2}\frac{\varphi }{2} &= \frac{1}{{{\mu ^2}}}\left( {\wp \left( {{\xi ^0} + {\omega _2}} \right) - {x_3}} \right) , \\
{\left( {{\partial _0}\varphi } \right)^2} &= 4\left( {{x_1} - \wp \left( {{\xi ^0} + {\omega _2}} \right)} \right) .
\end{align}
The sign of the quantities ${\cos}\frac{\varphi }{2}$, ${\sin}\frac{\varphi }{2}$ and ${{\partial _0} \varphi }$ depends on whether $\varphi$ is an oscillating or rotating solution. Although these signs are not going to play a crucial role in the following, equation \eqref{eq:elliptic_solution_phi_0} implies
\begin{equation}
\sign \cos \frac{\varphi }{2} = +1 , \quad \sign \sin \frac{\varphi }{2} = {\left( { - 1} \right)^{\left\lfloor {\frac{{{\xi ^0}}}{{2{\omega _1}}}} \right\rfloor }} , \quad \sign
{\partial _0}\varphi  = {\left( { - 1} \right)^{\left\lfloor {\frac{{{\xi ^0}}}{{2{\omega _1}}} + \frac{1}{2}} \right\rfloor }}
\end{equation}
for oscillating solutions, and
\begin{align}
\sign \cos \frac{\varphi }{2} = {\left( { - 1} \right)^{\left\lfloor {\frac{{{\xi ^0}}}{{2{\omega _1}}}} \right\rfloor }} , \quad \sign \sin \frac{\varphi }{2} = {\left( { - 1} \right)^{\left\lfloor {\frac{{{\xi ^0}}}{{2{\omega _1}}} - \frac{1}{2}} \right\rfloor }} , \quad \sign {\partial _0}\varphi = +1
\end{align}
for the rotating ones with increasing $\varphi$.

The equation \eqref{eq:sg_backlund_1} contains only the derivative of $\tilde{\varphi}$ with respect to $\xi^1$ and simultaneously all other functions that appear depend solely on $\xi^0$. Therefore, it can be solved as an ordinary differential equation, substituting the undetermined constant of integration with an undetermined unknown function of $\xi^0$. The latter equation assumes the form
\begin{equation}
{\partial _1}\frac{{\tilde \varphi \left( {{\xi ^0},{\xi ^1}} \right)}}{2} = A\left( {{\xi ^0}} \right)\cos \frac{{\tilde \varphi \left( {{\xi ^0},{\xi ^1}} \right) - {\hat \varphi}\left( {{\xi ^0}} \right)}}{2} + B\left( {{\xi ^0}} \right) ,
\label{eq:sg_backlund_1_cos_form}
\end{equation}
where
\begin{align}
A\sin \frac{{{\hat \varphi}\left( {{\xi ^0}} \right)}}{2} &=  - \frac{\mu }{2}\left( {a + {a^{ - 1}}} \right)\cos \frac{\varphi }{2} , \label{eq:ell_sin_phi0}\\
A\cos \frac{{{\hat \varphi}\left( {{\xi ^0}} \right)}}{2} &= \frac{\mu }{2}\left( {a - {a^{ - 1}}} \right)\sin \frac{\varphi }{2} , \label{eq:ell_cos_phi0}\\
B\left( {{\xi ^0}} \right) &=  - {\partial _0}\frac{\varphi }{2} .\label{eq:ell_B}
\end{align}
One should be careful in the inversion of \eqref{eq:ell_sin_phi0} and \eqref{eq:ell_cos_phi0}, so that $\hat \varphi$ is continuous and smooth and $A$ has the correct sign. Defining the inverse tangent function so that $\arctan x \in \left( - \pi / 2 , \pi / 2 \right)$, an appropriate selection for $\hat \varphi$ and $A$ is
\begin{align}
\hat \varphi &= 2\arctan \left( {\frac{{a - {a^{ - 1}}}}{{a + {a^{ - 1}}}}\tan \frac{\varphi }{2}} \right) + \left( {2k - 1} \right)\pi  + {\mathop{\rm sgn}} \left( {{a^2} - 1} \right)2\pi \left\lfloor {\frac{\varphi }{{2\pi }} + \frac{1}{2}} \right\rfloor , \label{eq:ell_tan_phi0} \\
A &= s_c \frac{\mu}{2} \sqrt {{a^2} + {a^{ - 2}} + 2\cos \varphi } , \label{eq:ell_A}
\end{align}
where $k \in \mathbb{Z}$ and we defined the sign $s_c$ as
\begin{equation}
s_c : = {{\left( { - 1} \right)}^k}{\mathop{\rm sgn}} a .
\end{equation}
For a translationally invariant oscillating seed solution given by \eqref{eq:elliptic_solution_phi_0} it holds that $\left\lfloor {\frac{\varphi }{{2\pi }} + \frac{1}{2}} \right\rfloor = 0$, whereas for a rotating one $\left\lfloor {\frac{\varphi }{{2\pi }} + \frac{1}{2}} \right\rfloor = \left\lfloor {\frac{{{\xi ^0}}}{{2{\omega _1}}} + \frac{1}{2}} \right\rfloor $.

Notice also that the monotonicity of $\hat \varphi$ is the same as that of the seed solution $\varphi$ when $\left| a \right| > 1$ and opposite when $\left| a \right| < 1$. We define
\begin{equation}
s_d : = \sign \left( \left| a \right| - 1 \right) .
\end{equation}

The quantity ${A^2} - {B^2} \equiv D^2$, which is going to play an important role in the following, is actually a constant, namely,
\begin{equation}
D^2 \equiv {A^2} - {B^2} = \frac{1}{4}\left[ {{\mu ^2}{{\left( {a - {a^{ - 1}}} \right)}^2} + 2\left({{\mu ^2} - E}\right) }\right] = \frac{1}{4} \left[ {{\mu ^2}{{\left( {a^2 + {a^{ - 2}}} \right)}} - 2 E } \right] .
\label{eq:SG_D2_def}
\end{equation}
For a given value of $E$, the constant $D^2$ may assume any value larger or equal to $D_{\min}^2 = \left( \mu^2 - E \right) / 2$. The latter assumes any given value larger than the minimum one, for exactly four distinct values of the \Backlund transformation parameter $a$; let $a$ be one of them, then the other three are $-a$ and $\pm 1/a$. Therefore, there is exactly one value of the \Backlund parameter $a$ corresponding to a given value of $D^2$ in each of the segments $\left( - \infty , -1 \right]$, $\left[ - 1 , 0 \right)$, $\left( 0 , 1 \right]$ and $\left[ 1 , \infty \right)$. There is an exception to this rule; there are only two distinct values of $a$, corresponding to the minimum possible value of $D^2 = D_{\min}^2$, namely $a = \pm 1 $.

It is clear that in the case of oscillating solutions, since $E < \mu^2$, the quantity ${D^2}$ is always positive. On the contrary, in the case of rotating solutions the sign of this quantity depends on the value of $a$. Therefore, for cases where ${D^2}$ can become negative we are able to select the sign of $A \pm B$, choosing the direction of rotation of the solution $\varphi$. In the following, we will assume that rotating solutions are characterized by increasing $\varphi$ and thus, for these solutions $B$ is always negative. We define
\begin{equation}
D := \begin{cases} \sqrt {{{A^2} - {B^2}}  } , & {A^2} - {B^2} > 0\\
- i \sqrt {{{B^2} - {A^2}}} , & {A^2} - {B^2} < 0 .
\end{cases}
\end{equation}
Substituting
\begin{equation}
\frac{{A + B}}{D}g = \tan \frac{{\tilde \varphi  - {\hat \varphi}}}{4} ,
\label{eq:elliptic_substitution}
\end{equation}
the equation \eqref{eq:sg_backlund_1_cos_form} assumes the form
\begin{equation}
\frac{{{\partial _1}g}}{{1 - {g^2}}} = \frac{D}{2} ,
\label{eq:elliptic_g_equation}
\end{equation}
whose solution is
\begin{equation}
g = \tanh \frac{D}{2}\left( {{\xi ^1} + f\left( {{\xi ^0}} \right)} \right) .
\end{equation}
Therefore, ${\tilde \varphi }$ assumes the form
\begin{equation}
{{\tilde \varphi } } = \hat \varphi  + 4\arctan \frac{{A + B}}{D}\tanh \frac{D}{2}\left( {{\xi ^1} + f\left( {{\xi ^0}} \right)} \right) .
\end{equation}

Returning to the \Backlund transformation \eqref{eq:sg_backlund_0} that we have not used so far, we may write it as
\begin{equation}
{\partial _0}\frac{\tilde \varphi }{2} = \frac{\mu }{2}\left( {a + {a^{ - 1}}} \right)\sin \frac{\varphi }{2}\cos \frac{{\tilde \varphi }}{2} - \frac{\mu }{2}\left( {a - {a^{ - 1}}} \right)\cos \frac{\varphi }{2}\sin \frac{{\tilde \varphi }}{2} ,
\end{equation}
since $\varphi$ does not depend on $\xi^1$. It is a matter of trivial algebra to write it in the form
\begin{multline}
{\partial _0}\frac{\tilde \varphi }{2} = \frac{\mu }{2}\left( {\left( {a + {a^{ - 1}}} \right)\sin \frac{\varphi }{2}\cos \frac{{{\hat \varphi}}}{2} - \left( {a - {a^{ - 1}}} \right)\cos \frac{\varphi }{2}\sin \frac{{{\hat \varphi}}}{2}} \right)\cos \frac{{\tilde \varphi  - {\hat \varphi}}}{2}\\
 - \frac{\mu }{2}\left( {\left( {a + {a^{ - 1}}} \right)\sin \frac{\varphi }{2}\sin \frac{{{\hat \varphi}}}{2} + \left( {a - {a^{ - 1}}} \right)\cos \frac{\varphi }{2}\cos \frac{{{\hat \varphi}}}{2}} \right)\sin \frac{{\tilde \varphi  - {\hat \varphi}}}{2} ,
\end{multline}
which is significantly simplified with the use of equations \eqref{eq:ell_sin_phi0} and \eqref{eq:ell_cos_phi0} to
\begin{equation}
{\partial _0}\frac{\tilde \varphi }{2} = \frac{{{\mu ^2}}}{{4A}}\left( {\left( {{a^2} - {a^{ - 2}}} \right)\cos \frac{{\tilde \varphi  - {\hat \varphi}}}{2} + 2\sin \varphi \sin \frac{{\tilde \varphi  - {\hat \varphi}}}{2}} \right) .
\end{equation}

Equation \eqref{eq:ell_tan_phi0} implies that ${\partial _0} {\hat \varphi} =  - {{{\mu ^2}\left( {{a^2} - {a^{ - 2}}} \right)B}}/({{2{A^2}}}) $, equation \eqref{eq:ell_A} implies that ${\partial _0}A = {{{\mu ^2}B}}\sin \varphi / ( {{2A}} )$, while equation \eqref{eq:ell_B} and the equation of motion imply that ${\partial _0}B = {{{\mu ^2}}}\sin \varphi / 2$. Finally, it holds that $ {\partial _0}g = {D}\left( {1 - {g^2}} \right)f'\left( {{\xi ^0}} \right) / 2 $. Performing the substitution \eqref{eq:elliptic_substitution} and putting everything together, we arrive at
\begin{equation}
f'\left( {{\xi ^0}} \right) = \frac{{{\mu ^2}\left( {{a^2} - {a^{ - 2}}} \right)}}{{4{A^2}}} = - \frac{{\frac{{{\mu ^2}}}{4}\left( {{a^2} - {a^{ - 2}}} \right)}}{{\wp \left( {{\xi ^0} + {\omega _2}} \right) - \frac{{{\mu ^2}}}{4}\left( {{a^2} + {a^{ - 2}}} \right) + \frac{{{E}}}{6}}} .
\end{equation}
The denominator in the above relation is always positive. Therefore, the sign of $f'\left( {{\xi ^0}} \right)$, and, thus, the monotonicity of $f\left( {{\xi ^0}} \right)$, is determined by the sign of the numerator. The function $f$ is increasing when $\left| a \right| > 1$ and decreasing when $\left| a \right| < 1$.

We define $\tilde{a}$ so that
\begin{equation}
\begin{split}
\wp \left( {\tilde a} \right) &=  - \frac{E}{6} + \frac{{{\mu ^2}}}{4}\left( {{a^2} + {a^{ - 2}}} \right) \\
&= x_1 + D^2 = {x_2} + \frac{{{\mu ^2}}}{4}{\left( {a - {a^{ - 1}}} \right)^2} = {x_3} + \frac{{{\mu ^2}}}{4}{\left( {a + {a^{ - 1}}} \right)^2} 
\end{split}
\label{eq:SG_patilde}
\end{equation}
and demand that it lies within the cell of the Weierstrass elliptic function defined by the four complex numbers $\pm \omega_1 \pm \omega_2$. 
The Weierstrass differential equation and equation \eqref{eq:SG_patilde} imply that $\wp {'^2}\left( {\tilde a} \right) = {{{\mu ^4}}}{D^2}{\left( {{a^2} - {a^{ - 2}}} \right)^2} / 4 $, which specifies $\tilde{a}$ up to an overall sign. We select the $\tilde{a}$ such that
\begin{equation}
\wp '\left( {\tilde a} \right) =
\frac{{{\mu ^2}}}{2}D\left( {{a^2} - {a^{ - 2}}} \right) 
\label{eq:SG_Pprimeatilde}
\end{equation}
or in other words, so that the real part of $\tilde{a}$ has always opposite sign than $s_d$.

Equation \eqref{eq:SG_patilde} implies that $\wp \left( {\tilde a} \right)$ is larger than at least two of the three roots. When $D^2 > 0$, it is also larger than the largest root, implying that $\tilde{a}$ lies in the real axis, in the interval $\left( 0 , \omega_1 \right)$, when $\left| a \right| < 1$, and in the interval $\left( -\omega_1 , 0 \right)$, when $\left| a \right| > 1$. When $D^2 < 0$, $\wp \left( {\tilde a} \right)$ lies between the two larger roots and therefore $\tilde{a}$ lies in the linear segment with endpoints $\omega_1 $ and $\omega_3 \equiv \omega_1 + \omega_2$, when $\left| a \right| > 1$, and $- \omega_1 $ and $- \omega_3 $, when $\left| a \right| < 1$. In the special limiting case $a = \pm 1$, the derivative of the function $f$ vanishes, and, thus, $\wp '\left( {\tilde a} \right)$ vanishes too. At this limit, $\wp \left( {\tilde a} \right)$ assumes the value of the root $x_2$, implying that $\tilde{a}$ is equal to $\pm \omega_1$ for oscillating backgrounds and $\pm \omega_3$ for the rotating ones. In the latter case, there is yet another $a$ for which $\tilde{a}$ assumes the value $\pm \omega_1$, and, thus, once again $\wp '\left( {\tilde a} \right)$ vanishes, which is the specific choice of $a$ that sets $D = 0$, namely, $a =  \pm \left( {E \pm \sqrt {E - {\mu ^2}} } \right)/\mu $.
\begin{figure}[ht]
\vspace{10pt}
\begin{center}
\begin{picture}(90,39)
\put(2,0){\includegraphics[width = 0.4\textwidth]{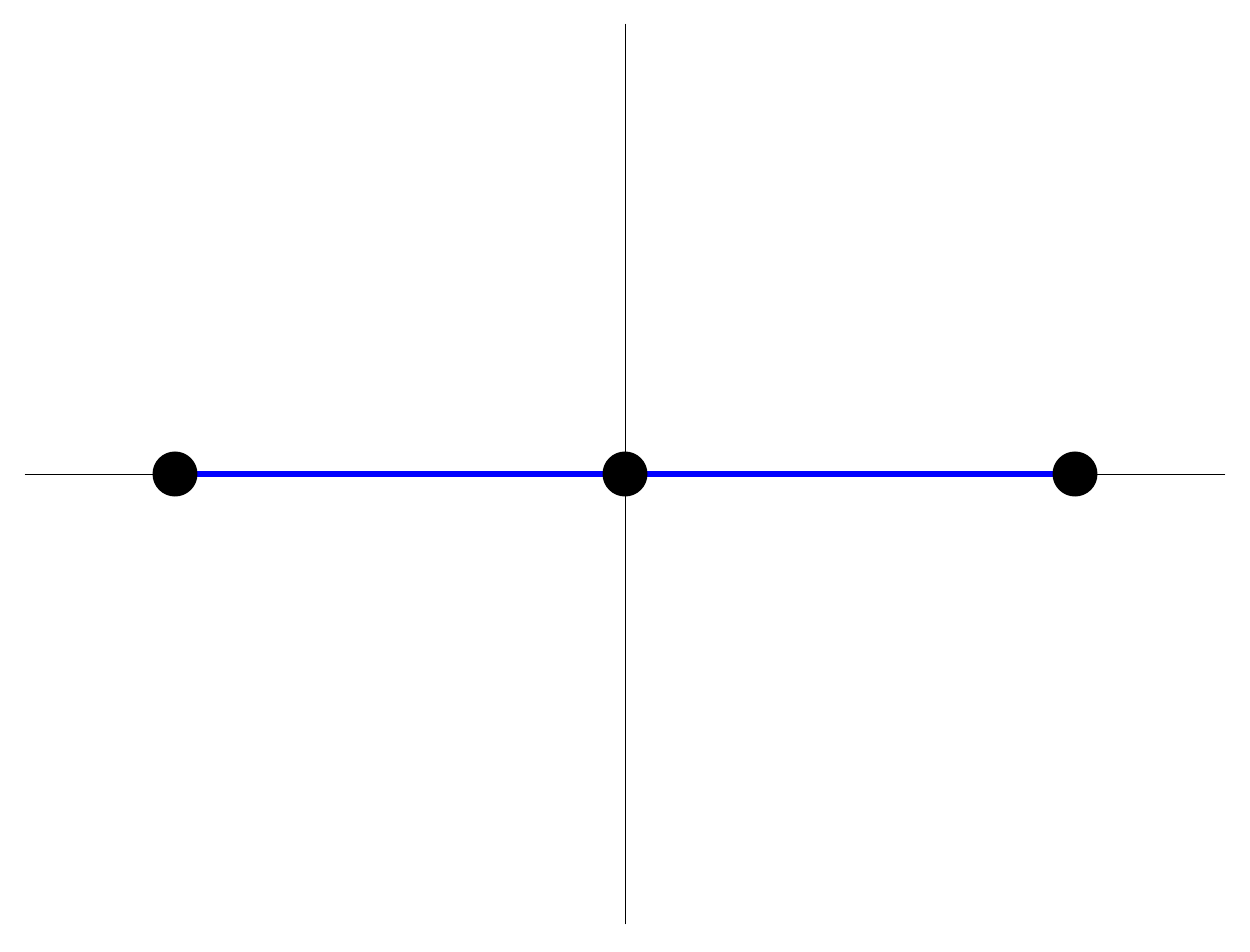}}
\put(45,0){\includegraphics[width = 0.4\textwidth]{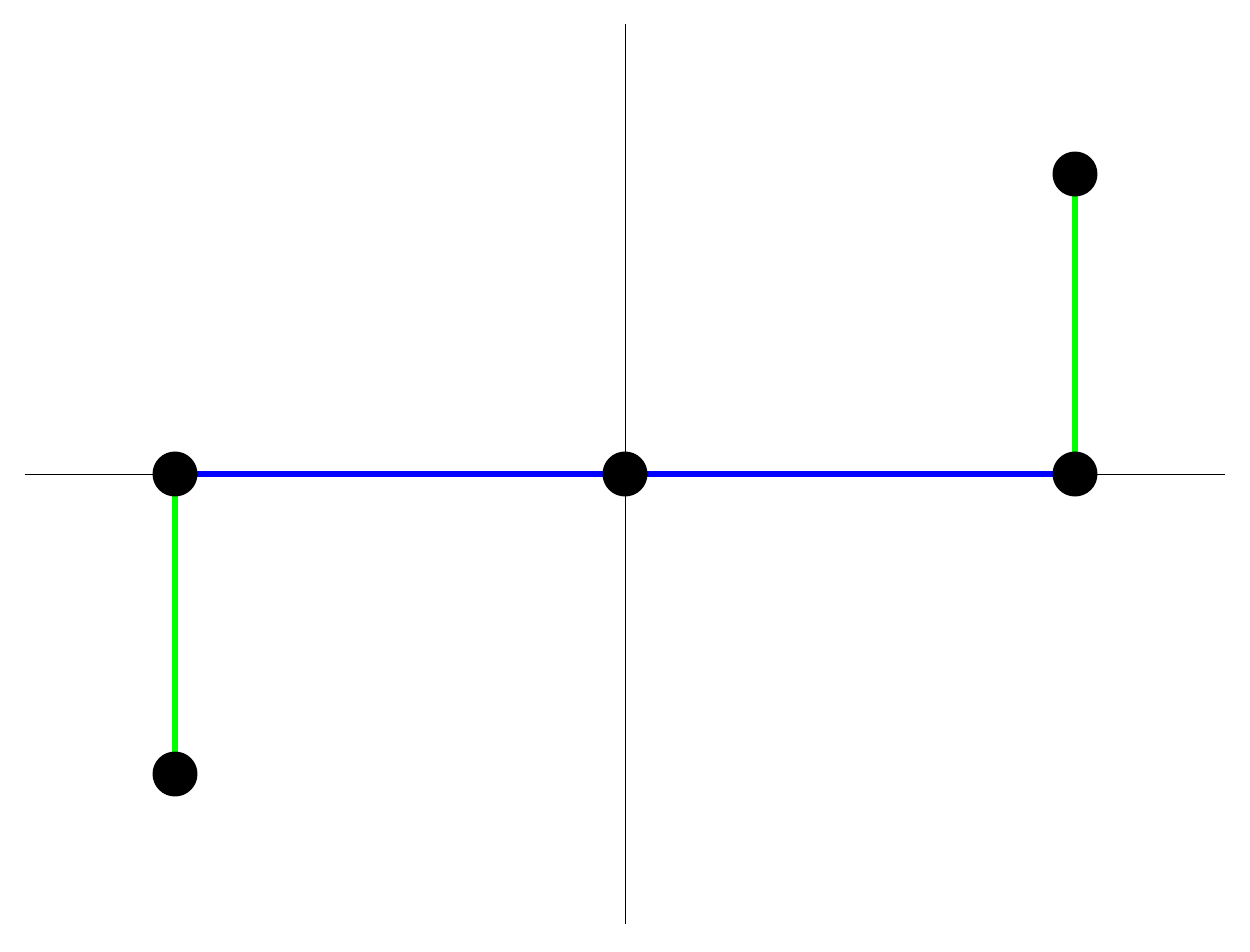}}
\put(75,10.5){$\tilde{a}=\omega_1$}
\put(74.25,7.5){$D^2=0$}
\put(73,33.5){$\tilde{a}=\omega_3$}
\put(66.75,30.5){$D^2=\left( \mu^2 - E \right) / 2$}
\put(72.75,27.5){$a=\pm 1$}
\put(54,20.25){$\tilde{a}=0$}
\put(50.25,17.25){$D^2 \to +\infty$}
\put(67.75,16){\color{blue}$D^2>0$}
\put(79.75,19.25){\color{green}$D^2<0$}
\put(32.25,10.25){$\tilde{a}=\omega_1$}
\put(26,7.25){$D^2=\left( \mu^2 - E \right) / 2$}
\put(32,4.25){$a=\pm 1$}
\put(11.25,20.25){$\tilde{a}=0$}
\put(7.5,17.25){$D^2 \to +\infty$}
\put(24.75,16){\color{blue}$D^2>0$}
\put(9.25,37){oscillating background}
\put(53.25,37){rotating background}
\put(74,6.75){\line(0,1){6.25}}
\put(74,6.75){\line(1,0){9.25}}
\put(83.25,6.75){\line(0,1){6.25}}
\put(74,13){\line(1,0){9.25}}
\put(82.75,13){\vector(-3,2){3}}
\put(66.5,26.75){\line(0,1){9.25}}
\put(66.5,26.75){\line(1,0){21.5}}
\put(88,26.75){\line(0,1){9.25}}
\put(66.5,36){\line(1,0){21.5}}
\put(83,26.75){\vector(-2,-1){3}}
\put(50,16.5){\line(0,1){6.25}}
\put(50,16.5){\line(1,0){13}}
\put(63,16.5){\line(0,1){6.25}}
\put(50,22.75){\line(1,0){13}}
\put(63,20.25){\vector(1,-3){1.5}}
\put(25.75,3.5){\line(0,1){9.25}}
\put(25.75,3.5){\line(1,0){21.5}}
\put(47.25,3.5){\line(0,1){9.25}}
\put(25.75,12.75){\line(1,0){21.5}}
\put(42.75,12.75){\vector(-3,1){6}}
\put(7.25,16.5){\line(0,1){6.25}}
\put(7.25,16.5){\line(1,0){13}}
\put(20.25,16.5){\line(0,1){6.25}}
\put(7.25,22.75){\line(1,0){13}}
\put(20.25,20.25){\vector(1,-3){1.5}}
\end{picture}
\end{center}
\vspace{-10pt}
\caption{The allowed values of $\tilde{a}$ in the complex plane. Each point in the $\tilde{a}$ complex plane corresponds to two discrete values of the \Backlund parameter $a$, differing only in their sign.}
\vspace{5pt}
\label{fig:atilda}
\end{figure}
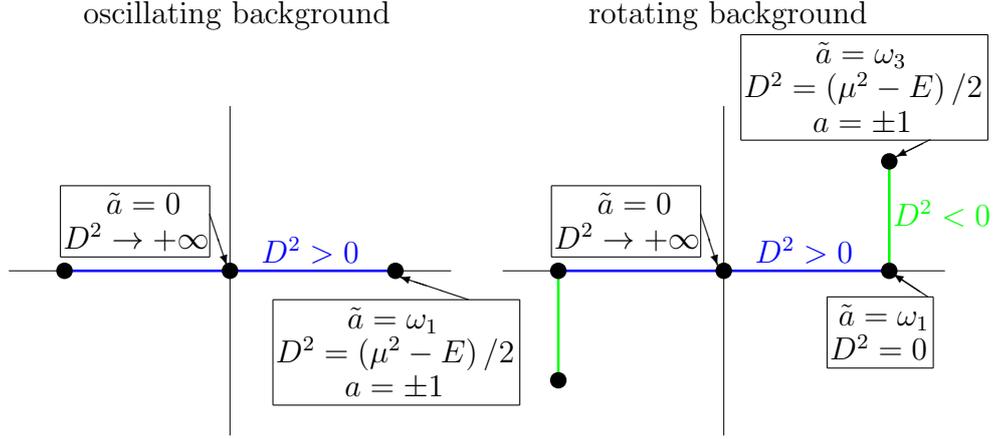

Using the above definitions, it can be shown that
\begin{equation}
f'\left( {{\xi ^0}} \right) =  - \frac{1}{{2D}}\frac{{\wp '\left( {\tilde a} \right)}}{{\wp \left( {{\xi ^0} + {\omega _2}} \right) - \wp \left( {\tilde a} \right)}}
\label{eq:elliptic_f_prime}
\end{equation}
implying
\begin{equation}
f\left( {{\xi ^0}} \right) =  - \frac{1}{{2D}}\left( {2\zeta \left( {\tilde a} \right){\xi ^0} + \ln \frac{{\sigma \left( {{\xi ^0} + {\omega _2} - \tilde a} \right)\sigma \left( {{\omega _2} + \tilde a} \right)}}{{\sigma \left( {{\xi ^0} + {\omega _2} + \tilde a} \right)\sigma \left( {{\omega _2} - \tilde a} \right)}}} \right) = \frac{i}{D}\Phi \left( {{\xi ^0};\tilde a} \right) ,
\label{eq:elliptic_f}
\end{equation}
where the function $\Phi$ is the same quasi-periodic function that appears in the expressions of the elliptic strings and it is defined in \eqref{eq:elliptic_solutions_lame_phase_def}. Putting everything together
\begin{equation}
\tilde \varphi  = {\hat \varphi} + 4\arctan \left[ \frac{{A + B}}{D}\tanh \frac{{D{\xi ^1} + i \Phi \left( {{\xi ^0};\tilde a} \right)}}{2} \right] .
\label{eq:SG_final_expression}
\end{equation}

Equations \eqref{eq:elliptic_f_prime} and \eqref{eq:elliptic_f} imply that when $D^2 > 0$, the function $\Phi \left( {{\xi ^0};\tilde a} \right)$ is purely imaginary, whereas when $D^2 < 0$, the function $\Phi \left( {{\xi ^0};\tilde a} \right)$ is real. Therefore, in all cases the solution $\tilde \varphi$ is real. It can be written in a manifestly real form as,
\begin{equation}
\tilde \varphi  = \begin{cases}
{\hat \varphi} + 4\arctan \left[ \frac{{A + B}}{D}\tanh \frac{{D{\xi ^1} + i \Phi \left( {{\xi ^0};\tilde a} \right)}}{2} \right] , & D^2 > 0 , \\
{\hat \varphi} + 4\arctan \left[ \frac{{ 1 - s_c}}{2} B\left( {{\xi ^1} + i \Phi \left( {{\xi ^0};\tilde a} \right)} \right) \right] , & D^2 = 0 , \\
{\hat \varphi} + 4\arctan \left[ \frac{{A + B}}{i D}\tan \frac{{i D{\xi ^1} - \Phi \left( {{\xi ^0};\tilde a} \right)}}{2} \right] , & D^2 < 0 .
\end{cases}
\label{eq:SG_final_expression_cases}
\end{equation}

Equation \eqref{eq:SG_final_expression_cases} reveals that there is a bifurcation of the qualitative characteristics of the dressed elliptic solutions of the sine-Gordon equation that occurs at $E=\mu^2$. As we have commented above, in the case of an oscillatory seed solution $D^2$ is always positive, whereas in the case of rotating seeds, there is a range of \Backlund parameters that sets it negative. Equation \eqref{eq:SG_final_expression_cases} implies that the solutions with $D^2>0$ look like a localized kink at the region $D{\xi ^1} + i \Phi \left( {{\xi ^0};\tilde a} \right) = 0$. Far from this region, they assume a form that is completely determined by the seed solution and it has the same periodicity properties as the latter. Thus, solutions with $D^2$ are localized disturbances on the elliptic background. On the contrary, solutions with $D^2<0$ do not have this property. They do not describe any kind of localized kink and they do not have the same periodicity properties as the seed solution in any region.

The same procedure can be repeated for a static elliptic seed solution. As expected by the symmetries of the sine-Gordon equation, the acquired solution reads
\begin{equation}
\tilde \varphi = {\hat \varphi} + 4\arctan \left[ \frac{{A + B}}{D}\tanh \frac{{D{\xi ^0} + i \Phi \left( {{\xi ^1};\tilde a} \right)}}{2} \right] ,
\label{eq:SG_final_expression_static}
\end{equation}
which can be acquired by equation \eqref{eq:SG_final_expression} interchanging the two coordinates and adding an overall angle $\pi$.

To sum up, the dressed elliptic string solution \eqref{eq:dressed_solution_final} has a sine-Grodon counterpart that is given by the equation \eqref{eq:SG_final_expression_static}, where the \Backlund parameter is given by the equation \eqref{eq:dress_vs_Backlund}.

The parameters appearing in the dressed string solutions and the solutions of the sine-Gordon equation presented in this section are also connected. The function $\Delta \left( \lambda \right)$ for $\lambda = e^{i\theta_1}$, which is the case of interest, is real and assumes the value $\Delta = - \left( \mu^2 \left( a^2 + a^{-2} \right) - 2 E \right) / 4$, where $a$ is given by \eqref{eq:dress_vs_Backlund}. This is exactly equal to the opposite of the parameter $D^2$ defined in \eqref{eq:SG_D2_def} that appears in the dressed elliptic sine-Gordon solutions. This is in line with the form of the dressed string solution; whenever $D^2$ is positive and thus $\Delta$ is negative, the trigonometric functions that appear in the dressed string solution will actually be hyperbolic functions when expressed in a manifestly real form, a fact expected for solutions with a kink counterpart.

Similarly, the function $\tilde{a} \left( \lambda \right)$, which appears in the dressed elliptic string solutions, when $\lambda = e^{i\theta_1}$ assumes a given value so that $\wp \left( \tilde{a} \right) = - E / 6 + \mu^2 \left( a^2 + a^{-2} \right) / 4$ and $\wp ' \left( \tilde{a} \right) = - i \sqrt{\Delta} \mu^2 \left( a^2 - a^{-2} \right) / 2$. Comparing to the defining properties \eqref{eq:SG_patilde} and \eqref{eq:SG_Pprimeatilde} of the parameter $\tilde{a}$ of the corresponding sine-Gordon solutions, the two parameters coincide, as long as one defines $\sqrt{\Delta} = i \sqrt{-\Delta}$, whenever $\Delta < 0$.
\pagebreak

\setcounter{equation}{0}
\section{Discussion}
\label{sec:discussion}

We presented the construction of dressed elliptic strings propagating on $\mathbb{R}^t \times$S$^2$. These solutions correspond to genus two solutions of the sine-Gordon equation with one of the two holes of the relevant torus being degenerate. Arbitrary genus solutions of both the sine-Gordon and the non-linear sigma model equations are known in an abstract form\cite{Kotlarov:2014yxa,Kozel,Dorey:2006zj}. Our approach adds to the relevant literature, because the solutions are expressed in terms of simple elliptic and trigonometric/hyperbolic functions, whose properties and qualitative behaviour are much easier to study and understand. Alternatively, specific non-degenerate genus two solutions can be constructed in a completely different approach \cite{Klose:2008rx}; the Pohlmeyer counterpart of the latter are genus-two solutions of the sine-Gordon equation \cite{LAMB:1971zz} that can be constructed via separation of variables after the application of the Lamb ansatz.

The dressing of the elliptic solutions is presented in both the Pohlmeyer reduced theory and the non-linear sigma model. In the first case it corresponds to a single \Backlund transformation, whereas in the second case to the application of the simplest possible dressing factor. Especially the latter calculation is an original non-trivial application of the dressing method, since the seed solution \cite{dual_spikes,helical,part1} is neither a solution whose Pohlmeyer counterpart is the vacuum, nor connected to this via a finite number of \Backlund transformations, as in most cases presented in the literature \cite{Combes:1993rw,Spradlin:2006wk,Kalousios:2006xy,Jevicki:2007pk,Jevicki:2007aa}. The similarities between the two pictures, even at technical level, reveal the deep connection between the dressing method and the \Backlund transformations \cite{Pohl_avatars}.

Independently of the choice of the seed solution, the special case where the dressing factor has the minimal number of poles, namely two poles lying on the unit circle, the effect of the dressing transformation on the seed solution acquires a nice geometrical picture. The dressed string is drawn by an epicycle of given radius, whose center runs over the seed solution. This picture adds to the conceptual understanding of the action of the dressing transformation on a given solution. It would be interesting to find the equivalent geometrical picture in other systems, such as strings propagating on AdS or dS spaces \cite{dual_spikes,spiky_string,bakas_pastras} or minimal surfaces in hyperbolic spaces \cite{Pastras:2016vqu}, as well as in the case of more general dressing factors.

In this work, the general solution to the auxiliary system for an elliptic seed solution \eqref{eq:dressed_string_auxiliary_solution} is obtained. Although we apply the simplest dressing factor, more complicated ones can be used in a straightforward way, without the need of solving again any differential equations. These dressing factors would correspond to performing multiple \Backlund transformations to the seed solution of the sine-Gordon equation. The above fact is connected to the existence of the addition theorem \eqref{eq:SG_addition_theorem}, which allows the performance of multiple \Backlund transformations algebraically.

Studying dressed solutions emerging from a dressing factor with four poles presents a certain interest, as an extension of our results. In the standard analysis, where the seed is the vacuum, such solutions correspond to the non-trivial scattering of two kinks or even bound states of the latter, the so called breathers. However, since in our case the seed solution already contains a train of kinks (or kink-antikinks) such phenomena appear in the dressed solutions we have studied, without the need of a second \Backlund transformation. The non-trivial interaction of the kink induced by the dressing with the kinks forming the background can be studied in the solutions with $D^2>0$, whereas a qualitatively different picture is expected whenever $D^2<0$. The study of more complicated dressed solutions however, will contain the extra feature of the non-trivial interaction of the two kinks that are both induced by the dressing in the presence of the non-trivial background.

Further investigation on the physical properties of the dressed elliptic strings is also very interesting. An interesting feature of the elliptic string solutions is the fact that they have several singular points, which are spikes. These can be kinematically understood, as points of the string that travel at the speed of light \cite{GKP_string} due to initial conditions. As they cannot change velocity, no matter what the forces are which are exerted on them, they continue to exist indefinitely, as long as they do not interact with each other. In the already studied spiky string solutions \cite{dual_spikes,spiky_string,multi,helical,Kruczenski:2006pk}, which are elliptic, the spikes rotate around the sphere with the same angular velocity, and thus, they never interact. Interacting spikes emerge in higher genus solutions. The simplest possible examples of this kind, which allow the study of spike interactions, are those obtained in this work.

The elliptic strings, are also characterized by a constant angular opening between consecutive spikes. The latter is holographically mapped to a quasi-momentum in the spin chain of the boundary theory. The dressed elliptic string solutions are not characterized by a single period, and, thus, their dispersion relations will depend on more than one quasi-momenta. Thus, these solutions may provide a tool for a further non-trivial check of the connection between the string dispersion relation and the anomalous dimensions of gauge theory operators in the strong coupling limit.

\subsection*{Acknowledgements}
The research of G.P. is funded by the ``Post-doctoral researchers support'' action of the operational programme ``human resources development, education and long life learning, 2014-2020'', with priority axes 6, 8 and 9, implemented by the Greek State Scholarship Foundation and co-funded by the European Social Fund - ESF and National Resources of Greece.

The authors would like to thank M. Axenides, E. Floratos, G. Georgiou and G. Linardopoulos for useful discussions.

\appendix

\renewcommand{\thesection}{\Alph{section}}
\renewcommand{\thesubsection}{\Alph{section}.\arabic{subsection}}
\renewcommand{\theequation}{\Alph{section}.\arabic{equation}}

\setcounter{equation}{0}
\section{Some Further Details on the Dressing Method}
\label{sec:dressing_formalities}
\subsection{The General Solution for the Residues}

In this appendix, we review the basic results of \cite{Harnad:1983we,Pohl_avatars} considering the specification of the residues appearing in the expression for the dressing factor and hence the specification of the dressed non-linear $\sigma$-model solution.

We consider the general form for the dressing factor as given by equations \eqref{eq:dressing_review_dressing_factor_general}. 
Let us first consider the case of poles obeying $\lambda_i\neq \mu_j$ for any $i$, $j$. Since obviously $\chi(\lambda)\chi(\lambda)^{-1} = I$, it holds that
\begin{equation}
\sum_i\frac{Q_i}{\lambda-\lambda_i}+\sum_i\frac{R_i}{\lambda-\mu_i}+\sum_{i,j}\frac{Q_iR_j}{(\lambda-\lambda_i)(\lambda-\mu_j)}=0 .
\end{equation}
Taking the residues of the above expression at $\lambda_i$ and $\mu_j$ yields the following relations for the yet unspecified matrices $Q_i$ and $R_i$,
\begin{equation}
Q_i+\sum_j\frac{Q_iR_j}{\lambda_i-\mu_j}=0 ,\quad R_i+\sum_j\frac{Q_jR_i}{\mu_i-\lambda_j}=0 .
\label{eq:dressing_review_QReq1}
\end{equation}
If the pole $\lambda_k$ coincides with the pole $\mu_l$, then the product $\chi(\lambda)\chi(\lambda)^{-1}$ will have a second order pole whose coefficient should vanish separately. In this case, vanishing of the residues at $\lambda = \lambda_k = \mu_l$ implies
\begin{equation}
Q_k R_l = 0 , \quad Q_k+\sum_{j\neq l} \frac{Q_k R_j}{\lambda_k-\mu_j}=0 ,\quad R_l+\sum_{j\neq k}\frac{Q_jR_l}{\mu_l-\lambda_j}=0 .
\label{eq:dressing_review_QReq1multi}
\end{equation}
Furthermore, $\Psi'\left( \lambda \right) = \chi\left( \lambda \right) \Psi \left( \lambda \right)$ must satisfy the equations of the auxiliary system \eqref{eq:dressing_review_current_dressed}. Substituting the expressions \eqref{eq:dressing_review_dressing_factor_general} into the latter and taking the residues at the positions of the poles yields
\begin{align}
(1\pm\lambda_i)\partial_\pm Q_i\left(1+\sum_j\frac{R_j}{\lambda_i-\mu_j}\right)+Q_iJ_{\pm}\left(1+\sum_j\frac{R_j}{\lambda_i-\mu_j}\right) &=0 , \label{eq:dressing_review_QReq2} \\
-(1\pm\mu_i)\left(1+\sum_j\frac{Q_j}{\mu_i-\lambda_j}\right)\partial_\pm R_i+\left(1+\sum_j\frac{Q_j}{\mu_i-\lambda_j}\right)J_{\pm}R_i &=0 . \label{eq:dressing_review_QReq3}
\end{align}

Equations \eqref{eq:dressing_review_QReq1}, \eqref{eq:dressing_review_QReq1multi}, \eqref{eq:dressing_review_QReq2} and \eqref{eq:dressing_review_QReq3} suffice to determine the matrices $Q_i$ and $R_i$ \cite{Harnad:1983we}. They equal
\begin{equation}
{Q_j} = \sum\limits_i {{M_{ij}}} ,\quad {R_i} =  - \sum\limits_j {{M_{ij}}} ,
\end{equation}
where
\begin{equation}
{M_{ij}} = \Psi \left( {{\mu _i}} \right){h_i}{\gamma _{ij}}f_j^\dag {\Psi ^{ - 1}}\left( {{\lambda _j}} \right) .
\end{equation}
The matrix $\gamma$ is the inverse of the matrix $\Gamma$ with elements given by
\begin{equation}
{\Gamma _{ij}} = \begin{cases} {{f_i^\dag {\Psi ^{ - 1}}\left( {{\lambda _i}} \right)\Psi \left( {{\mu _j}} \right){h_j}}} / \left( {{{\lambda _i} - {\mu _j}}} \right) , & \lambda_i \neq \mu_j, \\
- f_i^\dag {\Psi ^{ - 1}}\left( {{\lambda _i}} \right)\Psi '\left( {{\lambda _i}} \right){h_j} + f_i^\dag C{h_j}, & \lambda_i = \mu_j .
\end{cases}
\end{equation}
The vectors $f_i$, $h_j$ are arbitrary constant complex vectors, which obey $f_i^\dag {h_j} = 0$ when $\lambda_i = \mu_j$ and $C$ is an arbitrary constant matrix.

\subsection{The Constraints in the Case of Two Poles}
Our case of interest includes only a pair of poles, lying in the unit circle and being complex conjugate to each other. As we discussed in section \ref{subsec:dressing_involutions}, the unitarity involution enforces the poles of $\chi(\lambda)^{-1}$ to lie in positions complex conjugate to those of the poles of $\chi(\lambda)$. Thus $\mu_1 = \bar{\lambda}_1 = \lambda_2$ and $\mu_2 = \bar{\lambda}_2 = \lambda_1$. It can be shown that there is a particular solution for the residues, where the elements of the matrix $\Gamma$ connecting coinciding poles of $\chi(\lambda)$ and $\chi(\lambda)^{-1}$ are vanishing \cite{Harnad:1983we,SaintAubin:1982wrd}, namely $\Gamma_{12} = \Gamma_{21} = 0$. Therefore, for this particular solution, the matrix $\Gamma$ is diagonal and its inverse is obviously $\gamma = \diag \left\{ 1 / \Gamma_{11} , 1 / \Gamma_{22} \right\}$. Furthermore, it holds that $f_1^\dag {h_2} = f_2^\dag {h_1} = 0$.

The unitarity involution implies that the residues of the dressing factor obey $R_i = Q_i^\dag$. It turns out that this implies that $f_1 = h_1$ and $f_2 = h_2$. The reality involution implies that $Q_2 = \bar{Q}_1$. This in turn implies that $f_2 = \bar{f}_1$. The above are sufficient to conclude that the dressing factor is of the form given by equations \eqref{eq:dress_review_dressing_factor_two_conjugate_poles} and \eqref{eq:dress_review_dressing_factor_two_conjugate_poles_projector}, where $p \equiv f_1$. The coset involution implies that the residues should obey $Q_2 = - \lambda_2^2 f' \theta Q_1 \theta f$. This implies that $f_2 = \theta f_1$ or else the complex vector $p$ should obey $\bar{p} = \theta p$. Finally, the constraint $f_1^\dag {h_2} = f_2^\dag {h_1} = 0$ implies that the vector $p$ should obey $p^T p =0$. This concludes the derivation of this particular solution for the dressing factor in the case of two poles, complex conjugate to each other, that are lying on the unit circle, which is used throughout section \ref{subsec:dressed_strings_two_poles}.

\setcounter{equation}{0}
\section{Double Root Limits of the Dressed SG Solutions}

The dressed solutions of the sine-Gordon equation \eqref{eq:SG_final_expression} reduce to simpler expressions in the special case of a double root of the corresponding Weierstrass elliptic function. This is physically expected, since in these limits, the seed solution is either the vacuum or the one-kink solution, implying that the corresponding dressed solution should coincide to the one-kink or two-kink solution, respectively.

In the following, without loss of generality, we assume $a>1$. The first case to consider is the limit $E \to - \mu^2$. In the case of translationally invariant backgrounds this limit corresponds to the vacuum background $\varphi = 0$, and, thus, our expressions should degenerate to the well-known expressions of single kinks of the sine-Gordon equation. Indeed in this limit, the two smaller roots coincide, and, thus, the imaginary period diverges, whereas the real period acquires the specific value
\begin{equation}
{\omega _1} = \frac{{\pi }}{2 \mu } .
\end{equation}
The parameter $D$ acquires the value
\begin{equation}
D = \frac{\mu }{2}\left( {a + {a^{ - 1}}} \right) .
\end{equation}
Finally, it is a matter of simple algebra to show that the solution itself acquires the usual expression
\begin{equation}
\tilde \varphi  = 4\arctan {e^{\mu \left( {\frac{{a + {a^{ - 1}}}}{2}{\xi ^1} - \frac{{a - {a^{ - 1}}}}{2}{\xi ^0}} \right)}} .
\end{equation}

In the case of static backgrounds, in the limit $E \to - \mu^2$, the background solution tends to the vacuum $\varphi = \pi$ and the dressed solutions tend to solutions evolving from one unstable vacuum to another.

Another interesting case is the limit $E \to \mu^2$. In the case of a static background, the seed is a single static kink. Therefore, we should expect that our solutions should degenerate to the usual two-kink solutions of the sine-Gordon equation in the frame where one of the two is stationary. In this case, the two larger roots coincide, and, thus, the real period diverges. The background solution is written as
\begin{equation}
\cos \varphi  = 1 - \frac{2}{{{{\cosh }^2}\mu {\xi ^1}}}
\end{equation}
or
\begin{equation}
\varphi  = 4\arctan {e^{\mu {\xi ^1}}} .
\end{equation}
The parameter $D$ assumes the value
\begin{equation}
D = \frac{\mu }{2}\left| {a - {a^{ - 1}}} \right| .
\end{equation}
The parameter $\tilde{a}$ equals
\begin{equation}
\sinh \mu \tilde a = \frac{2}{{a - {a^{ - 1}}}} .
\end{equation}
The solution degenerates to the form
\begin{equation}
\tan \frac{{\tilde \varphi }}{4} = \frac{a-1}{a+1} \frac{{{e^{\mu \left( {\frac{{a + {a^{ - 1}}}}{2}{\xi ^1} + \frac{{a - {a^{ - 1}}}}{2}{\xi ^0}} \right)}} - {e^{ - \mu {\xi ^1}}}}}{{1 + {e^{\mu \left( {\frac{{a + {a^{ - 1}} - 2}}{2}{\xi ^1} + \frac{{a - {a^{ - 1}}}}{2}{\xi ^0}} \right)}}}} ,
\end{equation}
which is indeed the form of the two-kink solution in the frame that one of those is stationary. It corresponds to the outcome of the addition formula \eqref{eq:SG_addition_theorem} with $\varphi = 0$, $a_1 = -1$ and $a_2 = a$.

\end{document}